# Orbital order phase transition in $Pr_{1-x}Ca_xMnO_3$ probed by photovoltaics

B. Kressdorf[1], T. Meyer[2], M. ten Brink[3,4], C. Seick[5], S. Melles[1], N. Ottinger[1], T. Titze[5], H. Meer[5], A. Weisser[5], J. Hoffmann[1], S. Mathias[5], H. Ulrichs[5], D. Steil[5], M. Seibt[2], P.E. Blöchl[3,4], C. Jooss[1]

[1] *University of Göttingen, Institute of Materials Physics,*
[2] *University of Göttingen, 4th Institute of Physics,*
[3] *Clausthal University of Technology, Institute of Theoretical,*
[4] *University of Göttingen, Institute of Theoretical Physics,*
[5] *University of Göttingen, 1st Institute of Physics*

The phase diagram of $Pr_{1-x}Ca_xMnO_3$ is modified for $x \leq 0.3$, which suggests a reevaluation of the phase diagram of other manganites in that doping region. Rather than an orbital ordered phase reaching up to high temperatures of approximately 800-1100 K, we propose a loss of spontaneous orbital order already near room temperature. Above this temperature, the phase is characterized by a finite orbital polarization and octahedral tilt pattern. The tilt pattern couples to the Jahn-Teller distortion and thus induces a remaining orbital order, which persists up to high temperatures, where the tilt order is lost as well. This explains the experimental observation of orbital order up to high temperatures. The revaluation of the orbital order transition is based on observed anomalies of various physical properties at temperatures of 220-260 K in epitaxial thin films of $Pr_{1-x}Ca_xMnO_3$ $x=0.1$, i.e., in the photovoltaic effect, electric transport, magnetization, optical and ultrafast transient pump probe studies. Finite-temperature simulations based on a tight-binding model with carefully adjusted parameters from first-principles calculations exhibit an orbital-order phase transition at $T_{OO} \approx 300$ K for $Pr_{1-x}Ca_xMnO_3$ $x=0.1$. This is consistent with the experimental observation of a temperature dependent change in lattice parameter for bulk samples of the same doping at 300 K for $x=0.1$ and 350 K for $x=0$, typical for a second order phase transition. Since our reassignment of the orbital-order phase transition towards lower temperature challenges a well-established and long-accepted picture, we provide results of multiple complementary measurements as well as a detailed discussion.

## I. Introduction

Perovskite manganites are a class of materials with strongly correlated electronic, spin and phonon degrees of freedom. These correlations result in a complex phase diagram and a variety of exciting phenomena such as the colossal magnetoresistance (CMR) effect [1–3] and a photovoltaic effect [4,5]. As a consequence of the Jahn-Teller effect, $Mn^{3+}$ ions in manganites experience a strong electron-phonon coupling, which leads to the formation of polarons. The density of these polarons, and their electronic coupling can be controlled by composition. Depending on the density, a variety of ordered and disordered polaron arrangements can be formed. Among the manganites, the class of $Pr_{1-x}Ca_xMnO_3$ (PCMO) stands out because it allows to explore the effect of doping, while maintaining a similar electronic coupling (i.e. Mn-O-Mn hopping) [6]. Furthermore, the electron-phonon coupling in this class of manganites is strong compared to the electron hopping, which results in small, and thus stable, polarons and insulating ground states for the entire doping range [7].

At the heart of the manganite physics is the phase diagram and the nature of its individual phases. In this paper, we provide evidence that the phase diagram for PCMO accepted so far [8] may need to be reconsidered (see Fig. 1). In the low doping regime with PCMO x=0-0.2 a phase transition at 900 K for x=0 and at 675 K for x=0.1 [8] has been included, which is attributed to the order-disorder transition in the pattern of Jahn-Teller distortions. At higher doping, i.e. PCMO x=0.3, orbital and charge order exists only at a lower temperature up to $T_{CO}$=240 K. Our experimental and theoretical results indicate that in the low doping regime, spontaneous orbital order breaks down at temperatures much lower than previously believed. This finding at low temperatures requires a reinterpretation of the high-temperature regime of the phase diagram.

In the highly correlated PCMO the onset of a photovoltaic effect in the visible to infrared spectral range can be linked to the emergence of long living hot-polarons [9,10]. Heterojunctions composed of Nb-doped $SrTiO_3$ and PCMO with x=0.34 and x=0.95 show such a photovoltaic effect only below their respective ordering temperatures, i. e. in the charge and orbital ordered and in the antiferromagnetically ordered phases (Fig. 1 (b)). The connection of the photovoltaic effect to charge and orbital order has recently been demonstrated clearly by the observation of a pronounced photovoltaic effect at room temperature in the Ruddlesden-Popper manganite $Pr_{0.5}Ca_{1.5}MnO_4$ [11], which has a charge-order transition temperature of $T_{CO} \approx 320$ K. This raised the onset temperature of the photovoltaic effect above that of the parent compound PCMO at x=0.34. In order to extend the photovoltaic effect to even higher temperatures, we studied lightly doped PCMO with its high orbital ordering temperature. However, for PCMO x=0.1 we observe an onset of the photovoltaic effect only well below room temperature (Fig. 1 (b)). This discrepancy between the onset of the photovoltaic effect and the orbital ordering temperature inspired a detailed study of the structural and electronic properties of the temperature dependent orbital ordered state at low doping presented in this article.

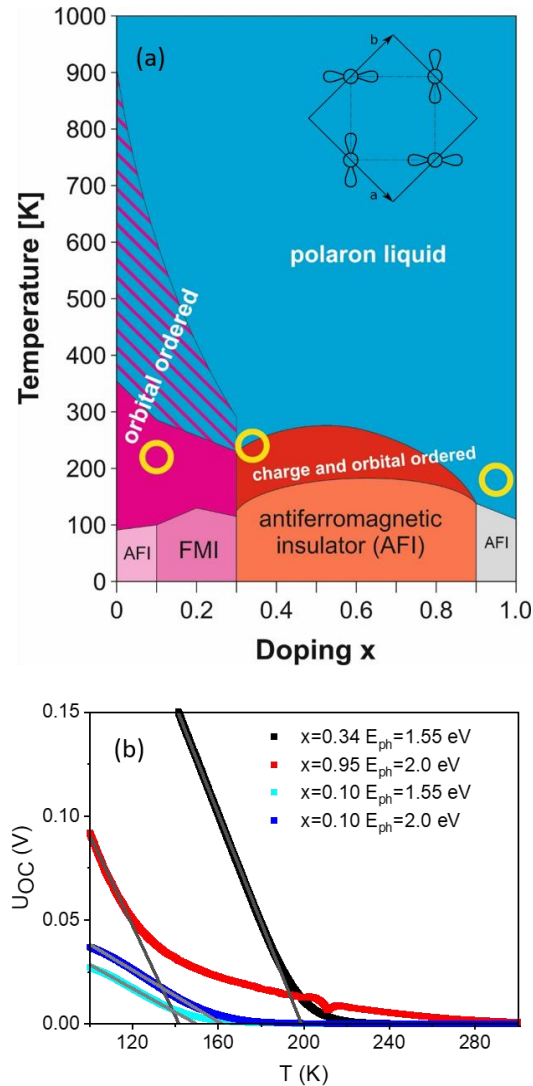

**FIG. 1:** Ordered states and onset of polaron photovoltaic effect in $Pr_{1-x}Ca_xMnO_3$ (PCMO). (a) Phase diagram in zero magnetic field according to [8,12,13]. Our results on the onset of polaron photovoltaic effect are indicated by yellow circles. Inset: Schematic sketch of orbital ordered structure of alternating orbitals for PMO. (b) Temperature

dependence of the open circuit voltage $U_{oc}(T)$ in $Pr_{1-x}Ca_xMnO_3$ – $SrTi_{1-y}Nb_yO_3$ heterojunctions for x=0.1, x=0.34 and x=0.95 for monochromatic illumination. The onset of linear increase coincides with the antiferromagnetic (x=0.95), charge / orbital order phase transitions (x=0.34) [10]. As it is shown in this article, the onset for PCMO x=0.1 indicates the orbital order phase transition.

Orbital ordered phases evolve in transition metal oxides where the degeneracy of d orbitals is lifted by the Jahn-Teller effect. In manganite perovskites, the Mn 3d-orbitals are split into three $t_{2g}$ states of $d_{xy}$, $d_{xz}$, and $d_{yz}$ symmetry and two $e_g$ states of $d_{x^2-y^2}$ and $d_{3z^2-r^2}$ symmetry due to the octahedral ligand field of the $MnO_6$ octahedra. At suitable doping with Mn valence close to $Mn^{3+}$, the degeneracy of these states is lifted by the Jahn-Teller distortion of the $MnO_6$ octahedra that leads to a preferential occupation of $d_{3x^2-r^2}$ or $d_{3y^2-r^2}$ states with one electron [14]. In an orbital ordered state, the two orbitals are alternately aligned, leading to an antiferro-type of orbital order with an order parameter $C_Q$. Such type of ordering was first discussed by Goodenough [15] in connection with experimental observations of structural distortions in $La_{1-x}Ca_xMnO_3$ [16]. Since then, orbital order was found for many hole doped manganites with small tolerance factor, the latter is an indicator for the stability and octahedral tilt distortion of the crystal structure based on ratio of the ionic radii. In such manganites, orbital order is present in certain doping regions in correlation with charge ordering, such as in $Pr_{1-x}Ca_xMnO_3$ (x=0.3-0.7) [8,17] and $Nd_{1-x}Sr_xMnO_3$ [18], in the un-doped $LaMnO_3$ [19] and $PrMnO_3$ [13], as well as in other type of transition metal oxides [20]. The orbital order ground state is well studied in theory [21,22].

In order to establish and study orbital ordered states, different methods have been applied. In half doped low bandwidth manganites, orbital ordering is accompanied by charge order. Here, the ordering of orbitals via ordering of formal $Mn^{3+}/Mn^{4+}$ valence states generates a super lattice that can be detected by diffraction experiments [23,24]. In fact, because the charge disproportion is rather small and typically amounts to only ~0.1e [25,26], the observed charge order/orbital order super lattice reflections mainly correspond to periodically varying Jahn-Teller distortions of the $MnO_6$ octahedra [24,27]. A powerful tool to directly study different occupation of the $d_{3x^2-r^2}$ and $d_{3y^2-r^2}$ orbitals are resonant scattering techniques at the Mn L and K edges with polarized x-rays, because they give direct access to the anisotropic charge density [28] and related orbital occupancy [29]. These resonant x-ray techniques are thus suitable to detect orbital ordered states at low doping levels, where charge order is absent. They have established ordering of $e_g$ orbitals in $LaMnO_3$ up to a temperature of 780 K, concomitant with a structural phase transition between the orthorhombic and pseudo-cubic phases [19]. In $PrMnO_3$, Sanchez et al. [30] showed evidence for the presence of orbital order via the structural refinement of distorted $MnO_6$ octahedra with three different Mn-O bonding length. The neutron diffraction results showed the presence of some degree of orbital order up to the temperature of the orthorhombic to pseudo-cubic phase transition at 1050 K referred to by the authors as a 'Jahn-Teller transition' [30].

The presence of some degree of ordering of Jahn-Teller distortions alone is, however, not sufficient to prove spontaneous orbital order. It is well known that in structural phase transitions, secondary order parameters can appear that break symmetries of the high temperature phase at the same transition temperature as the primary order parameter [31]. For example, in ferroelectric phase transitions in $BaTiO_3$ or $SrTiO_3$ the ferroelectric polarization is associated with the appearance of a tetragonal distortion, although the symmetry of both order parameters is different. In $SrTiO_3$, the ferroelectric state can be also induced by external strains, e.g. misfit strain in epitaxial films grown on different substrates [32]. In the case of orbital order transitions in $PrMnO_3$ as well as $LaMnO_3$, the onset of ordering of Jahn-Teller distortions coincidences with the tilting transition of $MnO_6$ octahedra which induces the transition between the cubic and orthorhombic phase. This phase transition is driven by the orthorhombic distortion that is related to steric effect and is controlled by the tolerance factor [33,34]. As shown by Alsonso et al. [35], the average tilt angle of $MnO_6$ octahedra increases with decreasing tolerance factor. Furthermore, the increase of octahedral tilt angle also slightly increases the amplitude of the Jahn-Teller distortion. Since at higher doping levels spontaneous orbital ordering sets in at quite low transition temperatures of the charge order / orbital order phase transition (e.g. PCMO x=0.5 $T_{CO}$=235 K [36] and LCMO x=0.5 $T_{CO}$=225 K [37]) this raises the question about the nature of the orbital order phase transition in the weakly-doped phases with PCMO x<0.3. In particular, the question needs to be considered, whether the transition to spontaneous orbital order appears below the pseudo cubic to orthorhombic phase transition and is partially hidden in the induced order of Jahn-Teller distortions driven by the tilting.

In this article, we show that orbital ordering in lightly doped $PrMnO_3$ most probably takes place near room temperature, i.e. at much lower temperatures than predicted by the phase diagram (Fig. 1 (a)). Notwithstanding the fact, that structural data alone only provide a subtle signature of this phase transition the presence of anomalies in electric transport, optical properties, magnetization and transient behavior of optical excitations provides clear evidence for the presence of a phase transition. They point to the emergence of a spontaneous orbital order below $T_{OO}$~220 K in thin films, which differs from the induced order of Jahn-Teller distortions at higher temperatures. In their bulk material, detailed temperature dependent x-ray diffraction measurements reveal an ordering temperature of 350 K for PCMO x=0 and 300 K for PCMO x=0.1. The finite-temperature simulations of the orbital-order phase transition based on a tight-binding model with carefully adjusted parameters from first-principles calculations confirm the presence of a spontaneous orbital ordering below $T_{OO} \approx$ 300 K for PCMO x=0.1.

## II. Experimental Results

In the following subsection, we report experimental results on various physical properties that provide a clear evidence for the presence of an electronic phase transition in $Pr_{0.9}Ca_{0.1}MnO_3$ (PCMO x=0.1) thin films at a temperature of around $T_{OO} \approx$ 220 K. These physical properties include (i) photovoltaic response (Sec. II.1), (ii) transport properties (Sec. II.2), (iii) magnetic ordering (Sec. II.3), (iv) optical properties (Sec. II.4), and (v) transient optical transmission (Sec. II.5). Preliminarily, we assign this phase transition as orbital ordering and come back to a critical evaluation after presentation of our experimental data and the results of the theoretical simulations.

### II.1 Photovoltaic properties of PCMO/STNO heterojunctions

Fig. 2 shows the change of the photovoltaic properties of $Pr_{0.9}Ca_{0.1}MnO_3$ (PCMO x=0.1) - $SrTi_{0.995}Nb_{0.005}O_3$ (STNO) heterojunctions in the vicinity of the phase transition. The cross-plane current-voltage characteristics exhibits a diode-like rectifying behavior which is evident from the exponential current increase in forward voltage direction and a smaller current, with a weak voltage dependence, in backward direction (Fig. 2 (a)). This diode-like behavior is observed throughout the temperature range of room temperature down to 100 K, where the rectifying characteristics becomes more pronounced below the respective ordering temperature. Under illumination, the photovoltaic effect of these heterojunctions appears as an additional photocurrent. The photovoltaic response is characterized by the parameters short-circuit current density $J_{sc}$ (at U=0V) and the open-circuit voltage $U_{oc}$ (at $I$=0 A). Their temperature dependences are displayed in Fig. 2 (b).

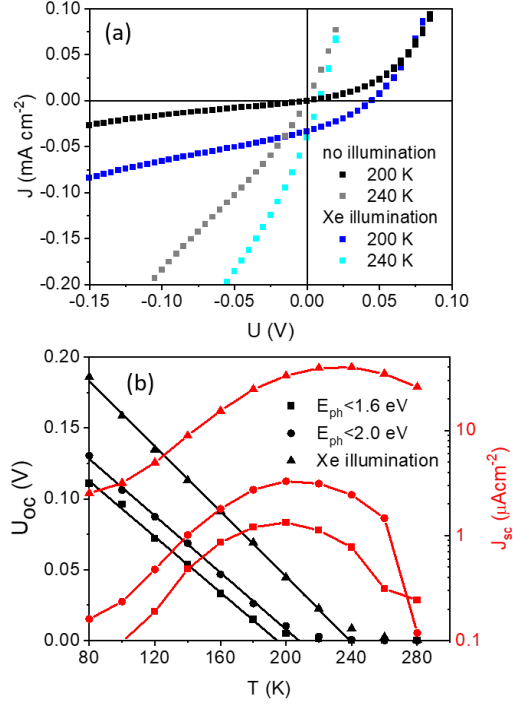

**FIG. 2**: Photovoltaic properties of the PCMO x=0.1 / STNO junction at the orbital order phase transition. (a) Current voltage characteristics above (T=240 K) and below (T= 200 K) the phase transition in the dark and under polychromatic illumination. (b) Temperature dependence of open-circuit voltage $U_{oc}$ and short-circuit current density $J_{sc}$ measured under different spectral illumination ranges: $E_{max}$=1.6 eV, $E_{max}$=2.0 eV and polychromatic illumination.

The photovoltaic effect under polychromatic illumination is measured using an UV-enhanced Xe lamp which produces a nearly constant spectral irradiance in the range between 1.55 and 4.1 eV. Additionally, measurements with two cut-off filters that limit the maximum excitation energy $E_{max}$ to 1.6 and 2.0 eV were realized. Both cut-off filters are well below the measured STNO bandgap of 3.3 eV [38], i.e. excess carrier excitations take place only in the manganite film.

Regardless of the selected type of illumination, the open-circuit voltage $U_{oc}$ and the short-circuit current density $J_{sc}$ show characteristic temperature dependences with two distinct temperature regimes: At high temperatures $U_{oc}$ is small, whereas, in the low temperature regime, a significant open-circuit voltage is observed, which increases nearly linear with decreasing temperature. In contrast, the temperature dependence of $J_{sc}$ exhibits a pronounced maximum in a temperature range of around 200 K (Fig. 2 (b)).

Interestingly, with an increasing spectral range of incident photons, $U_{oc}(T)$ is shifted to higher voltages. Since the open circuit voltage reflects the splitting of the chemical potential of electron and holes without carrier extraction, its increase with increasing $E_{max}$ mainly reflects the contribution of electron hole excitations at higher energy to $U_{oc}(T)$. Therefore, the "true" transition temperature $T_{OO}$~220 K is approximately presented by the $U_{oc}(T)$ onset at highest photon irradiance.

In the temperature range of the ordering transition, the short circuit current density $J_{sc}(T)$ exhibits a pronounced maximum (Fig. 2 (b)). Since optical absorption is more or less unchanged with temperature, the increase of $J_{sc}(T)$ with decreasing $T$ reflects an increase of the photo carrier lifetime when approaching the ordering transition. Below the phase transition, the carrier mobility decreases, as shown in the subsequent section. Similar to the shift of $U_{oc}(T)$ with increasing spectral illumination, also the maximum of the $J_{sc}(T)$ is shifted to higher $T$ with an increase of the current density. The $U_{oc}(T)$ onset and the maximum of $J_{sc}(T)$ are more or less at the same temperature. Such a behavior is unconventional and cannot be understood without the presence of changes in the electronic properties of the junction due to a phase transition in the temperature range between 190 K to 240 K.

## II.2 Charge carrier mobility

The change of electronic properties at the ordering transition is additionally reflected in a change of dc carrier transport of PCMO. The charge carriers in Ca-doped PrMnO$_3$ are small polarons that reveal a thermally activated mobility above half of the Debye temperature ½ $\theta_{Debye}$≈160 K [39]. In addition, a band-like transport can appear at low temperatures in high magnetic fields [7], via a magnetic field induced phase transition to a ferromagnetic metallic phase that evolves out of the charge ordered state of small polarons [40]. In a broad doping range of PCMO the temperature dependence of the resistivity can be described by thermally activated hopping of small polarons that is in the adiabatic approximation given by [41]:

(1a) $\quad \rho(T) = \rho_0 \, T \, e^{\frac{E_A}{kT}}$.

Electronic phase transitions commonly change the prefactor $\rho_0$ as well as the activation energy $E_A$. It is, therefore, useful to consider the logarithmic derivative defined as

(1b) $\quad E_A = kT \, \frac{d}{dt} ln\left(\frac{\rho}{T}\right)$

as a temperature-dependent activation barrier.

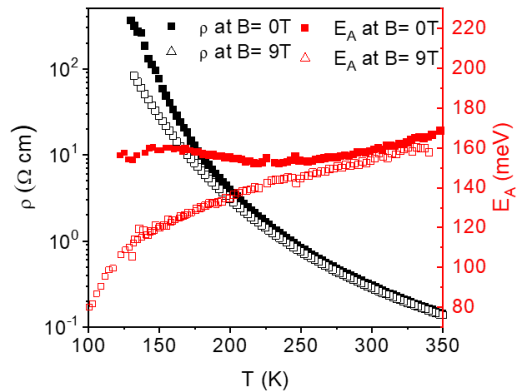

**FIG. 3**: Temperature dependence of resistivity and logarithmic derivative $E_A$ in zero magnetic field and at $\mu_0H$=9 T for a PCMO x=0.1 film on (001) STO.

Fig. 3 shows the increase of in-plane resistivity with decreasing temperature. The measurement is performed on a PCMO x=0.1 film deposited on insulating STO substrate that was post-annealed for 20 h at 900° C in order to reduce growth-induced defects and strain (see Supplemental Materials [42]). Above 220 K the resistivity shows an exponential temperature dependence that is reflected in a nearly constant logarithmic derivative with an average value of $E_A$≈160 meV. Below the ordering transition at $T_{OO}$≈220 K, $E_A(T)$ shows a significant increase, reflecting a change of the hopping mobility of small polarons in the low temperature state.

Interestingly, below the phase transition temperature, the resistivity shows a significant magneto-resistance effect. At an applied magnetic field of $\mu_0H$=9 T, the logarithmic derivative $E_A$ decreases almost linearly with decreasing temperatures down to about 150 K. Furthermore, the increase of $E_A$ below 220 K in zero-field is completely suppressed. The electric transport behavior supports the presence of orbital ordering below 220 K, since an ordering of the underlying Jahn-Teller distortions obviously reduces the mobility of Jahn-Teller polarons. As worked out by Millis, Littlewood & Shraiman [43], there is a strong mutual effect between the Jahn-Teller splitting of $e_g$ orbitals and the electronic hopping via the magnetic double exchange interaction. Our results indicate that increase of magnetic double exchange by a magnetic field induced spin alignment is more effective in the orbital ordered state. This is reflected in the observed magneto-resistance below $T_{OO}$.

## II.3 Magnetic properties

The observed magnetoresistance below $T_{OO}$ raises the question, whether the magnetic properties are affected by the ordering transition as well. According to the established bulk phase diagram of lightly doped PCMO in Fig. 1 (a), a magnetic phase transition takes place between a paramagnetic high-temperature phase and a magnetically ordered low-temperature phase. Its nature depends on the exact doping level since, at PCMO x=0.1, there is a phase boundary between a canted antiferromagnetic and a ferromagnetic phase [12,13]. The Neel and Curie temperatures of both magnetic phases are at about 100 K.

In thin film measurements, the contribution of substrate to the magnetic moment is usually not negligible. For STO, three relevant contributions need to be taken into consideration [44]: The intrinsic diamagnetic moment, a paramagnetic contribution due to impurities and a ferromagnetic surface magnetism $m_0$ that is closely related to the formation of oxygen vacancies. The surface contribution $m_0(H)$ is temperature independent and saturates at external field of about 200 mT.

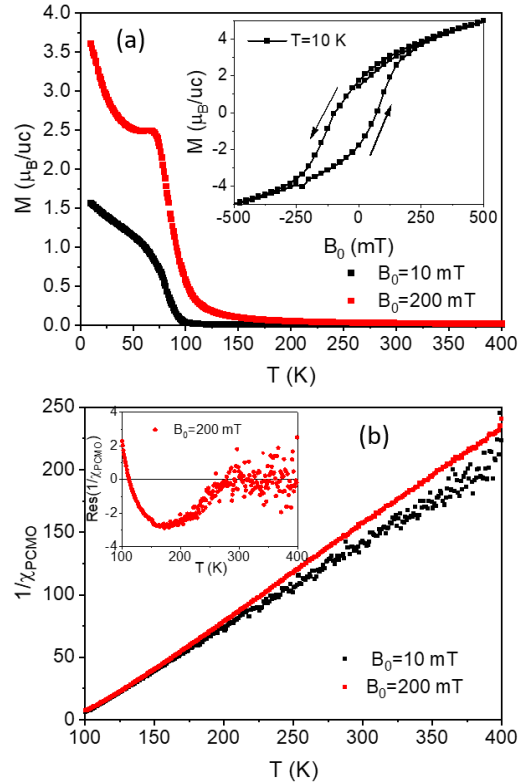

**FIG. 4:** Temperature dependence of field-cooled magnetization in Bohr magnetons per unit cell and magnetic susceptibility. The data is corrected with respect to substrate contributions, see text. (a) Temperature dependence of magnetization for different applied fields. The inset shows the hysteric behavior of field-dependent magnetization at 10 K. (b) Temperature dependence of the dc magnetic susceptibility at two different applied fields. The inset shows the residuum of the experimental data with respect to a linear fit at temperatures $T>300$ K

In order to separate substrate and film contributions, we first measured the field and temperature dependent moment $m_{STO}$ of a pristine STO substrate (see Supplemental Material [42]). In the next step, a PCMO x=0.1 film was deposited on this substrate and the same measurements of magnetic moment $m_{tot}(H, T)$ for film and substrate under identical measurement parameters were performed. It is known that in STO, a temperature independent ferromagnetic surface contribution $\Delta m_S$ can appear [44], in addition to the diamagnetic bulk contribution $m_{STO}(H,T)$. It depends on the oxygen vacancy concentration at the surface and, thus, it can change due to the heating of the substrate for thin film deposition. In order to take such changes into account and determine $\Delta m_S$, we have fitted the difference of the two experimental data sets $m_{tot}(H,T) - m_{STO}(H,T)$ in the high-temperature range ($T > 300$ K). Accordingly, the magnetic moment $m_{film}$ of the film then yields

(2) $\quad m_{film} = m_{tot}(H,T) - m_{STO}(H,T) - \Delta m_S(H) = \frac{\chi_0 H}{(T - T_{CW})}$

Here, $\chi_0$ represents the paramagnetic Curie constant and $T_{CW}$ the Curie-Weiss temperature. The correction $\Delta m_S$ is negligibly small (3 x $10^{-12}$ Am$^2$) for small fields of 10 mT but significant (1.6 x $10^{-9}$ Am$^2$) at 200 mT, where it is close to ferromagnetic saturation. Fig. 4 summarizes the magnetic properties of the PCMO x=0.1 thin film after the substrate correction.

At low temperatures, the field-cooled temperature and field dependent magnetization shows a distinct phase transition to a magnetically ordered state (Fig. 4 (a)). The strong increase of $M$ at $T_c$=80.3 ± 1.0 K is indicative for a transition into a phase with a ferromagnetic component, visible by the hysteretic behavior of $M(H)$ at 10 K and the presence of a remanent magnetization (see inset Fig. 4 (a)). Well above the Curie temperature $T_c$, the susceptibility of PCMO x=0.1 reveals Curie-Weiss behavior (Fig. 4 (b)). Linear fits in the temperature range of 120 K to 400 K yield a Curie constant that corresponds to a magnetic moment of 5.7 ± 0.3 $\mu_B$ and a Curie-Weiss temperature of $T_{CW}$=93.4 ± 0.2 K. The magnetic moment is close to the expected one, if the contribution of 4f moments of Pr is taken into account [45].

Interestingly, the slopes in the Curie-Weiss plot reveal slight differences in the high-temperature (*T*>280 K) and in the low-temperature (*T*<180 K) range. At higher temperatures, Fig. 4 (b) reveals a weak magnetic field dependency and the change in slope appears more pronounced for weak magnetic fields. Whereas the difference is close to noise for the measurement at 10 mT, the presence of two different temperature regimes is clearly visible in the residuum of the experimental data at 200 mT (inset Fig. 4 (b)). The deviation from the Curie-Weiss fit becomes significant below 260 K. This represents a probable magnetic fingerprint of a phase transition, in which the electronically ordered low temperature phase has a higher susceptibility. The error bars are too large for a quantitative comparison between orbital ordered and disordered phase, but the data suggests qualitatively that the Curie-Weiss temperature is lower and the Curie constant is higher in the orbital ordered phase. We can exclude that the change of χ(T) at 260 K is related to the transition into the ferromagnetically ordered state, since it the observed deviation from the Curie-Weiss fit is far above a possible onset of ferromagnetic fluctuations. This statement is supported by the additional change of the residuum (inset Fig 4 (b)) below a temperature of around 170 K that initiate the ferromagnetic transition below 100 K.

**II.4 Optical properties**

Fig. 5 shows the temperature dependence of the absorption coefficient $\alpha(E_{ph}) = 1 - t$, obtained through measurements of the transmission *t*, in the range of photon energies between 1.25 eV and 3.0 eV. It reveals 4 broad absorption bands at 1.8 eV (I), 2.4 eV (II), 2.7 eV (III) and above 3.2 eV (IV). The band IV here overlaps with the onset of interband transitions in SrTiO$_3$ with a bandgap of 3.2 eV. For $E_{ph}$≤3 eV, the absorption coefficient represents the manganite excitations only. The increase with decreasing temperature is most pronounced for the bands I and II, a behavior that is also observed for higher doping levels [6]. The reason is that peak I and II stem from transitions between Jahn-Teller split Mn3d $e_g$ states, i.e. involving Mn intersite transitions. Here, note that the nominal Mn-d $e_g$ states contain a contribution from the 'atomic' oxygen O2p states. In contrast, the absorption peaks above 2.2 eV (peak III and IV) are characterized as dipole allowed charge transfer transitions between O2p states and minority-spin $t_{2g}(\downarrow)$ and $e_g(\downarrow)$ states that are less affected by the orbital order phase transition [46].

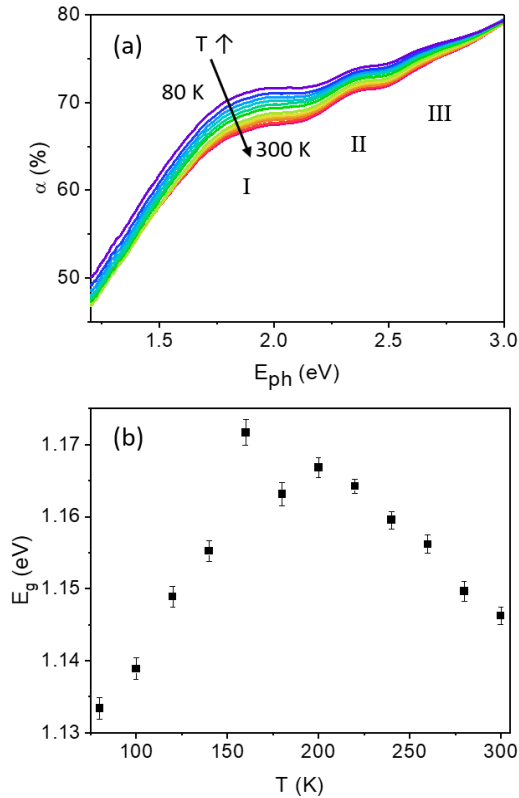

**FIG. 5:** Temperature dependence of optical absorption of PCMO x=0.1. (a) Spectral dependence of the absorption coefficient $\alpha$ ($E_{ph}$) at various temperatures. (b) Temperature dependence of the optical bandgap $E_g$.

In order to determine the bandgap, the temperature dependent reflection corrected absorption coefficient $\alpha_R(E_{ph})$ in the low photon energy region of the absorption band I was then analyzed by means of Tauc's relationship that is used to determine bandgaps in semiconductors [47]. The best fits were observed for a Tauc exponent of 3/2 that corresponds to direct forbidden transitions (see Supplemental Materials [42]). Since the transition I takes place between dipole allowed Jahn-Teller split occupied and unoccupied Mn 3d $e_g$-O 2p bands with different symmetry [6], the fit allows for a precise analysis of the bandgap. Even though we use the semiconductor theory, which is based on rigid bands, it is to be noted, that we have deviation of rigid bands due to the polaronic nature. The determined bandgap represents an upper limit to the real bandgap, since the transmission will be underestimated and the reflectance overestimated due to substrate contributions e.g. double reflection and interface contributions. However, this does not modify the temperature-dependent relative changes as well as peak positions.

Fig. 5 (b) shows the temperature dependence of the deduced bandgap $E_g$. The temperature dependence changes at about 200 K. At high temperatures we observe a bandgap opening with decreasing temperature. Such an effect is e.g. typically observed in semiconductors and has its origin in the thermal contraction as well as in the electron phonon interaction [48]. The latter smears out the bandgap at elevated temperatures. Remarkably, the bandgap decreases with decreasing temperature below the ordering transition. This is rather surprising, since an ordering in the orbital system decreases the thermal fluctuations of the Jahn-Teller split state and thus should increase the bandgap. A possible explanation may be the anomalous thermal expansion of PCMO x=0.1, where the c-axis expands with decreasing temperature [49].

**II.5 Time resolved optical transmission studies**

Optical pump-probe transmission measurements allow studying the change of the non-equilibrium quantities of the system, e.g., the lifetime of excitations, but also to access equilibrium properties such as the specific heat, which are otherwise difficult to measure in thin films. In our studies, a PCMO x=0.1 film (100 nm thick) is excited with a femtosecond laser pulse at fluences $F$ of 3-4 mJ/cm$^2$ with 1.2 eV pump photon energy, which is close to the bandgap. The time-dependent changes of the transmission were measured with a second pulse at 2.4 eV, which probes the band II, by dipole allowed intersite transitions between Jahn-Teller split occupied and unoccupied Mn 3d $e_g$ states [46]. Fig. 6 (a) shows a series of transient transmission time traces measured at cryostat temperatures between 20 K and 300 K. Three different timescales can be identified: an initial fast transient on the sub-ps timescale, a second transient on a timescale of about 10 ps, and long-term equilibration back to the ground state, which is not yet finished within the scan range of the pump-probe experiment. The transient transmission in Fig. 6 (a) additionally shows a short-lived oscillatory behavior $\Delta Tr$ in the first ps with a frequency of about 1.6 THz, which likely corresponds an optical phonon mode of the PCMO film.

Here, we are in particular interested in signatures of phase transitions, that manifest within transmission dynamics. We investigate two different quantities, which may contain information about a possible phase transition: First the signal decay $\tau_{Decay}$ back to the ground state, secondly the transmission change $\Delta Tr$ with temperature.

A phase transition between different ground states of the system will change the relaxation channels for the decay of the optically excited electron-hole pairs e.g. quasiparticle lifetimes and coupling constants between electrons, spins and lattice may change, as well as the heat capacities or thermal conductivities [50–53]. From Fig. 6 (a), we extracted the relaxation times after optical excitation back to the ground state (depicted in Fig. 6 (b) by fitting a simple exponential decay of the form

(3) $\quad \Delta Tr(t) = \Delta Tr(t < t_0) + A e^{-t/\tau_{Decay}}$

to the data starting from 90 ps delay. The extracted data in Fig. 6 (b)) shows an increase of the decay time from about 2 ns at room temperature to about 3.5 ns at low temperatures with a small maximum at about 200 K. This behavior is in stark contrast to the values computed from a simple three-temperature model shown in the (see Supplemental Materials [42]). Reflecting the strong increase of the thermal conductivity $\kappa$ of the MgO substrate with decreasing temperature from 52 W/mK at 300 K to 1850 W/mK at 20 K, the modelling predicts a significant decrease of the lifetime of the long-term decay of transient transmission $\Delta Tr$ from about 2 ns at 300 K to about 0.7 ns at 20 K. We note that this finding is very similar to results discussed in reference [9]. Similar to [9], this observation points towards the presence of electron-hole polaron excitations with rather long lifetime at low temperatures, which manifest as an additional signal contribution in the transmission change. This is supported by the observed photovoltaic response below the ordering transition.

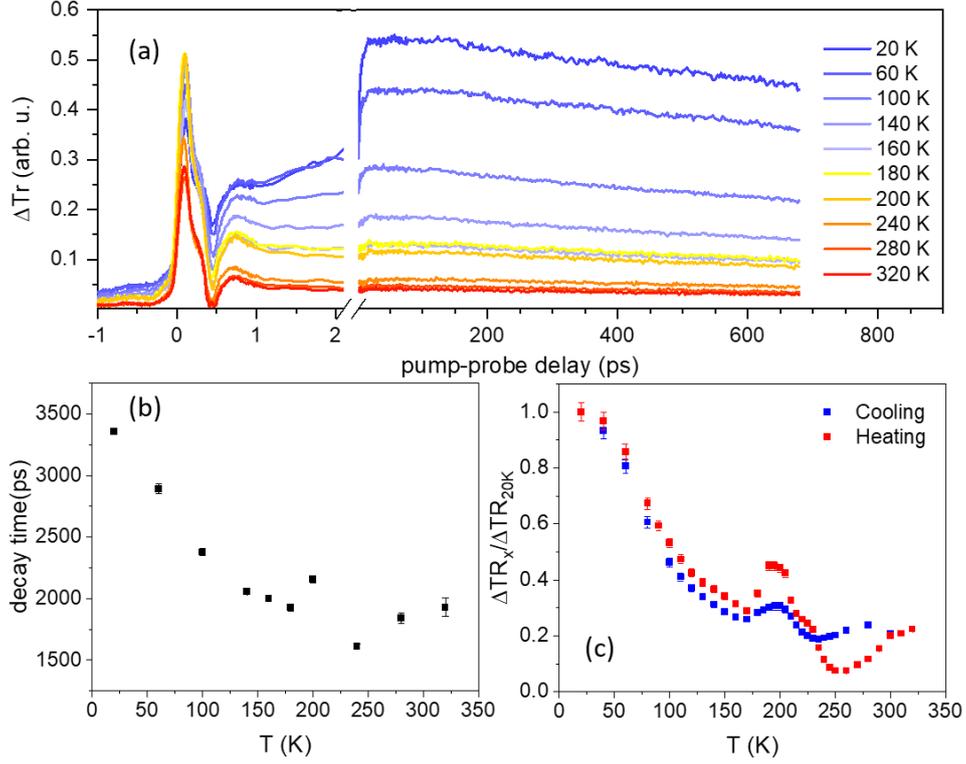

**FIG. 6:** Pump-probe transmission experiment for PCMO x=0.1 on MgO. (a) Transient transmission at different sample temperatures between 20 and 300 K for an incident fluence of 3.2 mJ/cm². (b) Change in the signal decay time back to the initial state with temperature showing a decrease with increasing temperature. A small anomaly is observed around 200 K. (c) Transient transmission change at about 90 ps delay for a combined cooling and heating temperature cycle with small temperature steps. The data is normalized to the value of the transient transmission change at 20 K. The strong anomaly observed for temperatures larger 170 K is due to a phase transition with hysteretic effects on temperature cycling.

Further information about a possible phase transition can be gained from inspecting an even simpler quantity from the data, the transmission change $\Delta Tr$ after initial equilibration of the optically excited electrons (Fig. 6 (c)). In the experiment, we maintain a constant incident optical fluence or energy density

(4a) $\quad F = \frac{E_{in}}{A}$

and vary the base temperatures $T_0$ inside the cryostat. Assuming purely thermal effects, for different $\Delta Tr(T_0)$ the observed transmission change is proportional to the change of the temperature $\Delta T$ caused by the transformation of the absorbed energy $E_{ab} \approx \alpha_{pump}(T_0) \cdot E_{in}$ into heat, where $\alpha_{pump}(T_0)$ is the absorption coefficient at the pump photon energy. For time scales larger than the initial relaxation time, the change in $\Delta Tr$ should then be mostly given by the temperature rise. This contribution can be approximated by

(4b) $\quad \Delta Tr \sim \frac{d\alpha_{probe\,eV}}{dT}\bigg|_{T=T_0} \Delta T = \frac{d\alpha_{probe}}{dT}\bigg|_{T=T_0} \cdot C(T_0)^{-1} \cdot E_{ab}$,

where,

(4c) $\quad \alpha_{probe}(T_0 + \Delta T) \approx \alpha_{probe}(T_0) + \frac{d\alpha_{probe\,eV}}{dT}\bigg|_{T=T_0} \Delta T$

is the temperature dependence of the absorption coefficient at the probe photon energy, and $C(T)$ is the specific heat. Due to the increase of $C(T_0)$ with increasing $T_0$, a continuous decrease in $\Delta Tr$ is expected for increasing the base temperature $T_0$ at constant incident optical fluence $F$. In the presence of phase transitions, the related non-monotonous change of $C(T)$ or other system quantities would then result in additional features.

Fig. 6 (c) shows the temperature dependence of $\Delta Tr$ for PCMO x=0.1 at about 90 ps delay time, i.e., after internal equilibration of the system is assumed for a combined cooling and heating cycle of the cryostat. Each depicted data point is the average of 300 measurements in the range 88-91 ps where the signal is nearly constant, error bars are the standard deviation. Below 170 K, $\Delta Tr$ indeed decreases with increasing base temperature as expected, most pronounced in the vicinity of the ferromagnetic phase transition temperature around 80 K. However, in the range starting from 170 K strong, step-like changes of $\Delta Tr$ are clearly visible, in addition to a significant hysteresis between the cooling and heating cycle. This is additional evidence for an orbital phase transition. Here, nonlinear behavior of material parameters like n, C, and κ, can give rise to a complex, non-monotonous behavior of the transmission change. The hysteresis in the cooling and heating cycle especially may indicate the coexistence of different phases in the discussed temperature range. In the simulation, a similar decrease of the signal can be seen (see the Supplemental Materials [42]), albeit it occurs significantly faster. This again shows that, in the experiment, additional physical processes - like a change of lifetime of polaronic excitations - are important. Note that the Supplemental Materials [42] contains the time-dependent evaluation of $\Delta Tr$ for all temperature-steps in Fig. 6 (a), showing the same kink in the data, but at much worse temperature-resolution for all times, showing that the result obtained here is not specific to 90 ps delay.

**II.6. Theoretical simulation of the orbital ordering transition**

**II.6.1 Model system**

The experimental results shown in the previous chapters clearly hint to an electronic phase transition in PCMO x=0.1 that appears at about 220 K in thin-film samples. In order to identify the nature of the phase transition, we performed finite-temperature simulations of the orbital order phase transition using a tight-binding model [54,55], which has been carefully adjusted to first-principles calculations [46].

The model captures the correlated motion of electrons, spins, and phonons. The Mn-$e_g$ electrons with two orbital degrees of freedom $j \in \{d_{3z^2-r^2}, d_{x^2-y^2}\}$ and two spin degrees of freedom $\sigma \in \{\uparrow, \downarrow\}$ per Mn-site are described by a Slater determinant of one-particle wave functions $|\Psi_n\rangle$. The half-filled shell of Mn-$t_{2g}$ electrons is accounted for by a spin $\vec{S}_R$ of length $\frac{3}{2}\hbar$. Two Jahn-Teller active octahedral distortions $Q_{2,R}$, $Q_{3,R}$ per Mn-site and one octahedral breathing mode $Q_{1,R}$ are considered [9,56]. The phonon amplitudes are extracted from the displacement of the oxygen ions along the Mn-O-Mn bridge, which accounts for their strongly cooperative nature. The oxygen atoms are limited to a one-dimensional motion along the oxygen bridge. We allow the lattice constants to adjust dynamically.

The Hamiltonian considers the intersite hopping of Mn-$e_g$ electrons and their onsite Coulomb interaction. The hopping parameters have been obtained by down folding the O-p orbitals of the oxygen bridge considering only $\sigma$-type matrix elements of the Mn-O axes. The spins of Mn-$e_g$ electrons experience a strong but finite Hund's coupling with the spins $\vec{S}_R$ of the $t_{2g}$ electrons. The latter experience an antiferromagnetic Heisenberg coupling between sites. A linear electron-phonon coupling [57] correlates the $e_g$ electrons with the Jahn-Teller active phonons $Q_2$ and $Q_3$ and the breathing mode $Q_1$. The Jahn-Teller active phonons lift the degeneracy of the $e_g$ electrons, and the outward breathing mode $Q_1$ stabilizes all $e_g$ electrons on this site. A detailed description of the model and its parameters can be found in an earlier publication [46].

The dynamics of the model is simulated in a Car-Parrinello framework [58]: The oxygen atoms are treated as classical particles that evolve according to Newton's equation of motion with forces obtained from the partial derivatives of the instantaneous total energy. Electrons and spins follow the atomic motion quasi-adiabatically in their instantaneous ground state. This quasi-adiabatic motion of electrons and spins is implemented by a fictitious Lagrangian. The temperature of the oxygen atoms is controlled by a Nosé-Hoover thermostat [59,60], which establishes a canonical (constant temperature) ensemble. More details of our simulations are given in Appendix B.

**II.6.2 Ground state**

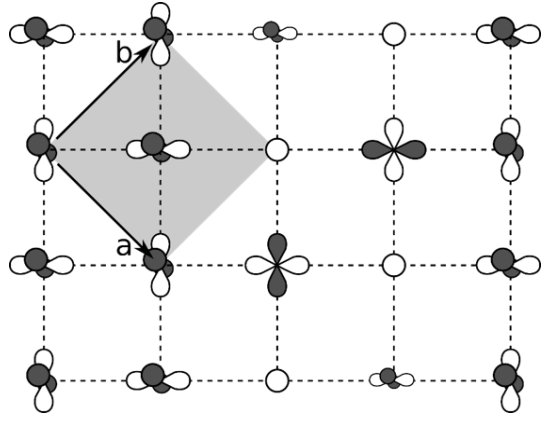

**FIG. 7**: Sketch of the orbital structures occurring in the low temperature phase at x≈0.1. We mainly observe the PrMnO$_3$ -like alternating orbitals $d_{z^2-x^2}$/$d_{z^2-y^2}$, together with clusters combining alternating hole and $d_{x^2-y^2}$ orbitals. Orbitals drawn with a smaller scale indicate a reduced orbital polarization, which can occur on sites between the two types of orbital ordering.

Below the orbital-order transition temperature, PCMO x=0.1 exhibits an orbital order similar to PrMnO$_3$, which is broken by the presence of hole polarons as sketched in Fig. 7. The PCMO x=0 orbital order consists of a checkerboard like arrangement of alternating $d_{z^2-x^2}$ and $d_{z^2-y^2}$ orbitals in the *ab*-plane.[1] The hole polarons are Mn ions in the formal 4+ oxidation state, i.e. they show a small charge deficiency and the absence of orbital polarization towards a certain $e_g$ orbital. The neighboring orbitals arrange preferentially with their lobes pointing towards the hole polarons. This leads to cluster formation of hole polarons and Mn-sites with in-plane $d_{x^2-y^2}$ orbitals. The formation of such clusters in neighboring *ab*-planes is unfavorable. At very low temperatures the hole polaron clusters are concentrated in single *ab*-planes. We found no influence of this ordering on the orbital order phase transition investigated in this paper. More disordered states with the hole polaron clusters scattered in many of the *ab* -planes have an only slightly increased ground state energy by a few meV.

The PMO -like orbital order prefers an antiferromagnetic order along the c-axis, while the hole polarons favor a ferromagnetic alignment in that direction. The competition of these effects can lead to spin-canting. A common problem of model Hamiltonians of the type used here is that they yield a *c/a* ratio larger than one, which differs from experiment [8]. We are not aware of a non-trivial cure for this problem.

**II.6.3 The order parameter**

The orbital order of the system is described by the structure factor

$$(5) \quad C_Q(\vec{q}) = \frac{1}{N_R} \sum_{j \in \{2,3\}} \left| \sum_R e^{-i\vec{q}\vec{R}_R} Q_{j,R} \right|^2$$

of the Jahn-Teller distortions $Q_{2,R}, Q_{3,R}$. $N_R$ is the number of Mn-sites in the sum. Due to the finite super-cell size in our calculations, this correlation function has contributions only at discrete points in reciprocal space. We define the order parameter for the orbital order phase transition as the structure factor summed along the $q_c$-direction:

$$(6) \quad C_Q^{av}(q_a, q_b) = \sum_{q_c} C_Q(q_a, q_b, q_c) \text{ at } (q_a, q_b) = \left(\frac{2\pi}{a}, 0\right)$$

in Pbnm notation. This order parameter quantifies the checkerboard like alternating orbital order in the ab-plane. In our simulations the orbital order of the Mn-$e_g$ electrons appears simultaneous with the ordering of the Jahn-Teller distortions. A correlation function of the Mn-$e_g$ electrons yields the same transition temperature as the Jahn-Teller order parameter used here.

---

[1] The orbital shape in the calculations differs slightly from $d_{3x^2-r^2}$ and $d_{3y^2-r^2}$ proposed on experimental grounds [74].

### II.6.4 Simulations

In finite-temperature simulations for PCMO x=0.1, the orbital order melts at 270 K. Figure 8 (a) shows the order parameter $C_Q^{av}\left(\frac{2\pi}{a}, 0\right)$ for the orbital order as a function of temperature. Both, the heating (red) and the cooling (blue and magenta) curves, are shown to ensure that the hysteresis due to the finite simulation time is negligible. As a result of the finite super-cell used in the calculations, the transition is not abrupt. The sharpest drop of the order parameter occurs around 270 K, where it passes its half maximum value. Above 400 K, the Jahn-Teller correlation function becomes nearly independent of the wave vector as seen in Fig. 8 (b). The correlation length of the antiferro-distorsive order reduces to nearest neighbors only at high temperatures,

The Jahn-Teller distortions are present over the entire temperature range investigated, i.e. T<800 K, but the correlation of the distortions at different sites vanishes above the transition temperature due to the breakdown of orbital order. In Fig. 8 (b), the *ab*-plane Jahn-Teller correlation function $C_Q^{av}(q_a, q_b)$ is shown for a typical heating cycle at 20 K, 280 K and 400 K. The 20 K and 400 K correlation functions are fully reproducible for all simulation runs. In the 280 K calculations we observed a few runs that have a more pronounced (0,0) peak. The correlation function for the finite super cell has contributions only at discrete points in reciprocal space, which have been broadened by a Gaussian in Fig. 7 (b). The mean intensity is the measure for the expectation value of the Jahn-Teller distortion.

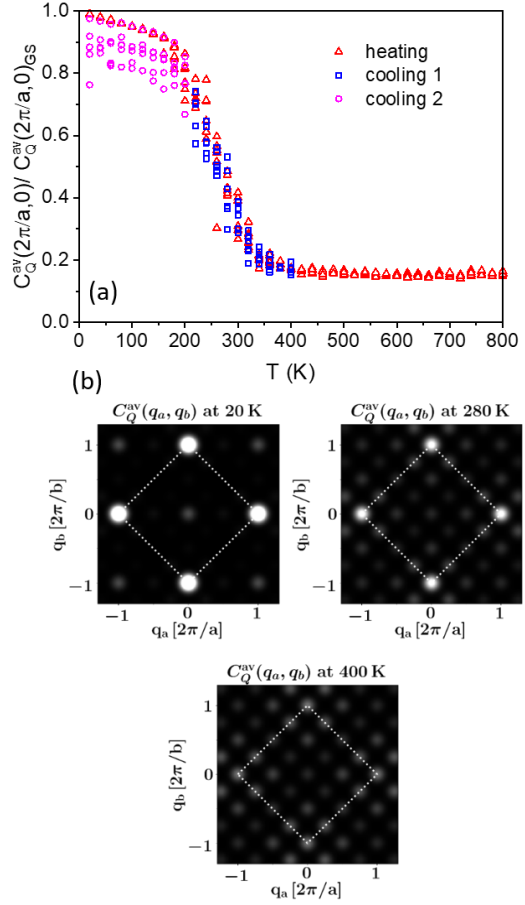

**FIG. 8**: (a) Order parameter $C_Q^{av}\left(\frac{2\pi}{a}, 0\right)$ for the orbital order transition in PCMO x=0.1 as function of temperature obtained in our simulations, scaled to its ground state value. Six simulation runs are shown for each temperature. The system was first heated from the ground state to non-zero temperatures, shown with red crosses. The system was subsequently cooled from the 400 K heating calculations to lower target temperatures (cooling 1), and in a second step from the 200 K cooling calculations back to 20 K (cooling 2), shown with blue squares and magenta circles, respectively. More details on the heating and cooling cycles are provided in the Supplemental Material at [42] (b) In-plane Jahn-Teller correlation function $C_Q^{av}(q_a, q_b)$ of a typical heating simulation close to the ground state (20 K), during the transition (280 K) and above the transition (400 K). The white dotted line marks the shape of the reciprocal unit cell. The peaks have been broadened by a Gaussian. Near the ground state we clearly see the PrMnO$_3$-like alternating orbital structure at $(q_a, q_b) = \left(\frac{2\pi}{a}, 0\right)$, resp. $(q_a, q_b) = \left(0, \frac{2\pi}{b}\right)$. At 400 K nearly all possible diffraction spots of our unit cell have the same weight.

These model simulations indicate the presence of an orbital order phase transition at around 270 K, close to the experimental values presented in this paper. A comment on the magnetic phase transition is provided in the Supplemental Materials [42].

### II.7. Temperature dependent structural changes in bulk PCMO

### II.7.1 Temperature dependent changes of lattice constants

The theoretical simulations give clear evidence for the presence of the orbital ordering phase transition in the vicinity of room temperature. This temperature is somewhat higher compared to the experiments in PCMO x=0.1 thin films where relevant physical properties changes in the vicinity temperature of 220-280 K. A reduced ordering temperature in the thin films compared to bulk is expected due to growth induced defects and substrate induced epitaxial strain.

After observing small changes or anomalies in various physical properties, the question is obvious whether such a phase transition might be visible in a change of the lattice parameters, as well. Therefore, we have performed temperature dependent XRD measurements on polycrystalline PCMO x=0 and PCMO x=0.1 powder samples in order to avoid substrate induced strain effects in the thin films and to get access to a higher number of different crystal orientations. The lattice parameters *a, b* and *c* were deduced from the (200), (020) and (004) reflections, respectively, and their temperature dependences are shown in Fig. 9. A careful inspection of the intensity distributions shows that the XRD reflections reveal a double-peak structure (Fig. 9 (a)) for both systems, PCMO x=0 and PCMO x=0.1. The shape of the doubled reflection peaks does not change with temperature. The double-peak structure originates from $K\alpha_1$ and $K\alpha_2$ irradiation contribution. With respect to temperature dependence, the lattice constants deduced from the $K\alpha_1$ peak may include a systematic but temperature independent error. The room temperature lattice parameters are in good agreement with experimental data from Refs [8,30,61].

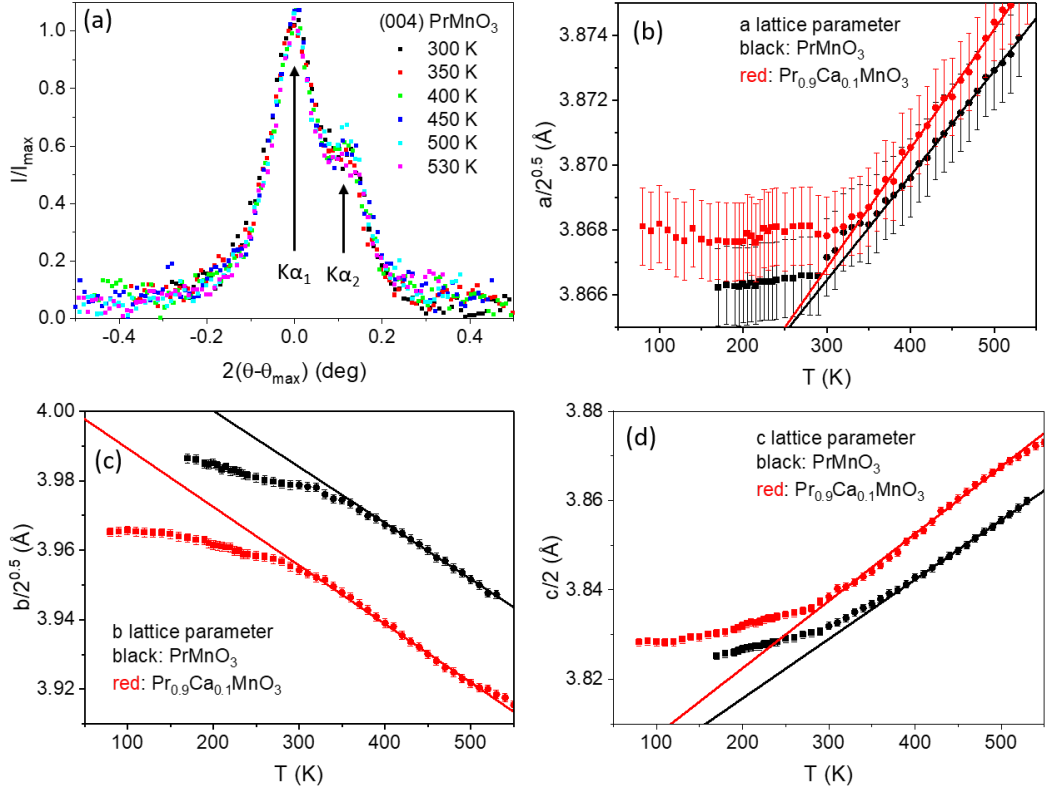

**FIG. 9:** (a) Intensity distributions of (004) PCMO x=0 reflections at different temperatures. The data is normalized to the position and the height of the intensity maximum. Temperature dependence of lattice parameters of PCMO x=0 and PCMO x=0.1 polycrystalline bulk samples (b-d). The lattice parameters correspond to Pbnm notation and are converted into a pseudocubic setting in order to facilitate the quantitative comparison.

The temperature dependence of the lattice parameters (Fig. 9 (b)-(d))) exhibits a contraction in the *c*-direction and an expansion in the *b*-direction while cooling. The *a* lattice parameter decreases as well, but the change is very small in comparison to the changes in *b* and *c*. These trends are typical for the orthorhombic *Pbnm* structure of PCMO and are fully consistent with the results of references [13,30] for PMO. However, our detailed temperature scan reveals the presence of two regimes with distinctly different behavior of the lattice parameters *a(T), b(T)* and *c(T)*. We observe a significantly reduced temperature dependence below 350 K for x=0 and 300 K for x=0.1 For the *a* lattice parameter this reduction even results in a temperature independent value within the experimental errors. The comparison of PCMO x=0 and PCMO x=0.1 as seen in Fig. 9 reveals a similar temperature dependence with comparable temperature slopes but a lower transition temperature for the lightly doped system. The shift of *a* and *c* to higher and of *b* to lower values for the doped system is as expected for a reduced orthorhombicity in the Ca-doped system compared to PMO.

In the temperature region below 150 K the lattice parameters for PCMO x=0.1 exhibit a second change in slope, where they tend to become temperature independent. This happens at temperatures far below the Debye temperature, where a saturation of thermal expansion is expected. The appearance of temperature independent lattice parameters may reflect the well-established second phase transition into magnetic ordered state [49].

Usually, thermal expansion coefficients reveal a similar temperature dependence as the specific heat [62], thus they strongly change below the Debye temperature, where the specific heat of materials is typically strongly temperature dependent. To the best of our knowledge, no measurements of the specific heat of PCMO x=0 and PCMO x=0.1 near room temperature are published. The decrease of the thermal expansion at 350 K, not far below the Debye temperature of about 460 K [63], thus represents an anomalous thermal expansion behavior and implies an anomaly of the thermal properties in this temperature regime.

The available data on temperature dependence of the specific heat of lightly Ag-doped PrMnO$_3$ (Pr$_{1-x}$Ag$_x$MnO$_3$, x ≤0.2) does not show any anomaly that would indicate at phase transition in the temperature range between 200 and 320 K [64]. This raises the question whether the orbital order transition, if present at lightly Ag doped PrMnO$_3$, is a continuous transition since the structural changes between a tilt induced order and spontaneous order are rather small.

### II.7.2 Change in lattice parameters for Jahn-Teller modes

In the following, we want to consider whether the changes of the lattice parameter can be interpreted as a structural fingerprint of Jahn-Teller distortions with respect to the Jahn-Teller modes $\Delta Q_2$ and $\Delta Q_3$. Therefore, we study the influence of Jahn-Teller distortions as well as octahedral tilt on the orthorhombic lattice parameters [65]. The atomic positions in bulk PCMO x=0 and PCMO x=0.1, obtained in Rietveld refinement by Jirak et al. [8,65] at room temperature provide our starting point. In accordance with Tamazyan et al. [66], the lattice parameters of a corner-shared network of MnO$_6$ octahedra can be expressed as a function of the tilt angle $\theta$ and the Jahn-Teller modes $\Delta Q_2$ and $\Delta Q_3$ as followed:

(7a) $\quad a = \sqrt{2}(d_l + d_s)\left(\cos\theta + n_1^2(1-\cos\theta)\right) + \sqrt{2}(d_l - d_s)(n_1 n_2(1-\cos\theta) - n_3 \sin\theta)$

(7b) $\quad b = \sqrt{2}(d_l + d_s)\left(\cos\theta + n_2^2(1-\cos\theta)\right) + \sqrt{2}(d_l - d_s)(n_2 n_1(1-\cos\theta) + n_3 \sin\theta)$

(7c) $\quad c = 4\, d_m\left(\cos\theta + n_3^2(1-\cos\theta)\right)$

Here, (n$_1$, n$_2$, n$_3$) is the normal vector along the tilt axis, and d$_l$, d$_m$, d$_s$ the long, medium, and short Mn-O distance of the MnO$_6$ octahedra. The octahedra are assumed to be ideal, i.e. no distortion of the right angles appears as justified by the work of Zhou and Goodenough [67]. Combined with values obtained by Jirak et al. [8,65] at room temperature, the changes in Mn-O bonding lengths can be parametrized by the relative changes of the amplitude of the two Jahn-Teller modes $\Delta Q_1$, $\Delta Q_2$ and $\Delta Q_3$ compared to their room temperature values:

(7d) $\quad d_l = d_l^{RT} + 1/\sqrt{3}\ \Delta Q_1 + 1/\sqrt{2}\ \Delta Q_2 - 1/\sqrt{6}\ \Delta Q_3$

(7e) $\quad d_m = d_m^{RT} + 1/\sqrt{3}\ \Delta Q_1 + 2/\sqrt{6}\ \Delta Q_3$

(7f) $\quad d_s = d_s^{RT} + 1/\sqrt{3}\ \Delta Q_1 - 1/\sqrt{2}\ \Delta Q_2 - 1/\sqrt{6}\ \Delta Q_3$

Here, we assume that the average Mn-O distance remains unchanged, i.e. no temperature dependent breathing of the octahedra are accounted for and thus $\Delta Q_1 = 0$. In a next step the direction of the rotation axis is set to the values given by the refinements [8,65] and it is assumed that the direction of the axis remains temperature independent. Subsequently, the root mean square difference between modelled and experimental lattice parameters is minimized by varying $\Delta Q_2$, $\Delta Q_3$, and tilt angle $\theta$. This yields temperature dependent values for the Jahn-Teller modes $\Delta Q_2$, $\Delta Q_3$ as well as the tilt angle, which are presented in Fig. 10 (a). Whilst the tilt angle remains relatively unchanged, the increased anisotropy in lattice parameters can be described with enhanced Jahn-Teller distortions towards low temperatures. The comparison between modelled and experimental lattice parameters is shown in Fig. 10 (b) and shows a very good quality of the fit.

Noteworthy, the fixation of the rotation axis as well as the mean Mn-O distance to the literature values at room temperature is a model assumption chosen to avoid an underdetermined minimization problem, possibly affecting the quantitative values of the extracted parameters. However, while the rotation axis is known to be rather insensitive to doping and temperature changes [8,30,65], a change in the mean Mn-O distance affects all lattice parameters, equally. A uniform outward/ inward breathing distortions $\Delta Q_1$ is thus unsuitable to explain the increased anisotropy at low temperatures. Consequently, the experimentally found data can be described by changes in the Jahn-Teller modes $\Delta Q_2$ and $\Delta Q_3$, only. The parallel trends of $\Delta Q_2$ and $\Delta Q_3$ indicate that a very similar type of continuous changes of the amplitudes of the Jahn-Teller distortions takes place in PCMO x=0 and PCMO x=0.1 that mainly increases the antiferrodistorsive in-plane ordering $\Delta Q_2$ with decreasing temperature.

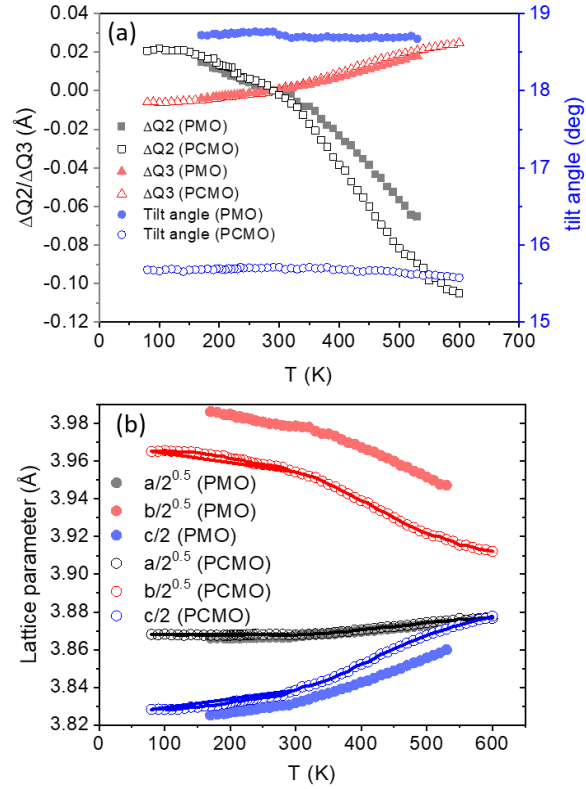

**FIG. 10:** (a) Change of Jahn-Teller modes $\Delta Q_3$ and $\Delta Q_2$ relative to the room temperature values as well as octahedral tilt angle for PCMO x=0 (closed symbols) and PCMO x=0.1 (open symbols). (b) Pseudo-cubic lattice parameters corresponding to the $\Delta Q_2$, $\Delta Q_3$ and tilt angle $\theta$ shown in (a) (symbols) as well as the experimentally obtained values replotted from Fig. 8 (lines).

### III. Discussion

The comprehensive temperature dependent study of physical and structural properties of PCMO x=0.1 presented above indicates the presence of a low temperature phase transition, which has not been reported in previous research of the phase diagram of PCMO. The main changes of the physical properties of the films observed at the orbital order phase transition at $T_{OO}\approx$200-250 K can be summarized as follows:

(1) The hot-polaron photovoltaic effect sets in at $T_{OO}$ (Fig. 2) and therefore depends on orbital-order (for PCMO x=0.1), respectively on orbital and charge order (for PCMO x = 0.34) [10]. An increased hot-polaron lifetime below $T_{OO}$ has been identified by optical pump probe experiments in (Fig. 6).

(2) A subtle but reliable change of the activation barrier for thermally activated hopping of hole-type polarons is observed at $T_{OO}$ (Fig. 3). Such an increase of the activation barrier with decreasing temperatures can be explained through the formation energy of orbital defects in an orbital ordered structure. Such an excess energy is not required in an orbital disordered structure, or is lower in a structure with a lower degree of order.

(3) Another fingerprint of the phase transition is the observed onset of a pronounced magnetoresistance effect below $T_{OO}$. Furthermore, a slight change in the temperature dependence of the magnetic susceptibility χ(T) is visible. The onset of this change in χ(T) is observed at 260 K, which is above, but close to $T_{OO}$. Typically, phase transitions that affect spin properties already manifest themselves at higher temperatures in the magnetic susceptibility through a change in spin fluctuations. The physical origin might be a change in the double exchange coupling between Mn ions due to an increase in orbital order, as e.g. described in the model of Millis et al. [43].

(4) The temperature dependence of the optical bandgap shows a transition from a normal behavior with an increase with decreasing temperature due to thermal contraction to an anomalous behavior below $T_{OO}$ (Fig. 5), where the bandgap decreases during cooling. This may reflect the anomalous thermal expansion behavior below the transition temperature, where the thermal expansion coefficients change due to the ordering. The change in optical properties is also reflected in a change of the transient optical response after pumping polaron excitations at the bandgap: Polaronic excitations do not only show an increase in the lifetime below $T_{OO}$, the transient absorption also exhibits non-monotonous and hysteretic features, pointing to an anomaly in the specific heat in this temperature range.

(5) The lattice parameters of two bulk powder PCMO samples with doping of x=0 and x=0.1 exhibit a change in thermal expansion at around 350 K and 300 K, respectively, which we attribute to the orbital-order transition. This gives us access to the orbital-order temperature in the bulk as opposed to films, for which the orbital order transition temperature is lower due to misfit strain and growth-induced defects.

**Orbital order and orbital polarization**

Our theoretical calculation of the orbital-order phase transition, exhibits an onset of spontaneous orbital order near 270 K for PCMO x=0.1 and preliminary results for PCMO x=0 yield a transition temperature of 360 K. At the transition temperature, the calculated order parameter for orbital order $C_Q^{av}$ for $\vec{q} = \left(\frac{2\pi}{a}, 0\right)$ falls off rapidly with increasing temperature. The orbital polarization on the other hand, is not affected by the orbital-order transition and remains intact up to high temperatures. While the **orbital polarization** is a measure of the size of the distortion in a single octahedron, the **orbital order** describes the correlation of distortions on different octahedra.

The orbital polarization is quantified by the corresponding structural parameter, the relative distortion $D = \frac{1}{6}\sum_{j=1...6}[(d_j - \langle d \rangle)/\langle d \rangle]^2$ with the Mn-O distances $d_j$ in the octahedron and their mean value $\langle d \rangle = \frac{1}{6}\sum_{j=1,..,6} d_j$. The relative distortion is directly related to the integral $D = \frac{1}{12\langle d\rangle^2}\int d^3q\, C_Q(\vec{q})$ of the correlation function $C_Q$, when only the Jahn-Teller active distortions $Q_2$ and $Q_3$ are considered.

Specifically, the correlation functions exhibit peaks, which broaden as the systems loses the orbital order.[2] The broadening lowers the peak intensity at the wave vectors of the lattice, but not its integrated weight. The width is a measure of the (inverse) correlation length and thus of orbital disorder. The total weight of the peaks, on the other hand, is a measure of orbital polarization.

We note, that our simulations do not incorporate the tilt order of the material. Therefore, the remaining induced orbital order observed experimentally [19] above the orbital order transition, which we attribute to an orbital order induced by the octahedral tilt, is absent.

**The nature of non-magnetic thermal phase transitions at low doping**

Our results indicate that the orbital order transition in the low-doped region occurs at considerably lower temperatures than previously believed. For bulk PCMO x=0 our results indicate the transition to be near 360 K, which is consistent with the simulations. This conclusion affects also the assignment of the high-temperature phase transition above 800 K [13,30,68] for PCMO x=0, previously assigned to be due to orbital order. At this point, however, we can only speculate about the nature of the phases in the high-temperature part of the phase diagram:

At high temperatures, Pollert et al. [13] determined two phase transitions in PCMO x=0, namely an orthorhombic to orthorhombic O/O' transition at 815 K and the orthorhombic to pseudocubic O'/C transition at 945 K. Sanchez et al. [30] obtained a similar pattern, albeit with transitions shifted to higher temperatures, namely with the O/O' transition at 948K and the O'/C transition at $T_{JT}$=1050 K.[3] The O'/C transition is related to a strong increase of the electric conductivity with increasing temperature [68,69]. In the orthorhombic low-temperature phase, conductivity is restricted due to thermally activated polarons, while metallic band conduction dominates in the high-temperature phase. Zhou and Goodenough [68] also describe a transition temperature $T^*<T_{JT}$ which is characterized by a slope-discontinuity in the thermoelectric power. It has to remain open at this point whether $T^*$ can be identified with the O/O' transition observed by Pollert et al. [13].

The loss of orbital order alone, does not explain the transition to metallic conduction, because it preserves the Jahn-Teller splitting between the $e_g$ orbitals. This notion is supported by our finding in the simulations, which exhibit a finite bandgap also above the orbital ordering transition.

A possible mechanism for the metal-insulator transition is as follows: A reduction of the mean Mn-O-Mn bond angle increases the band width, which in turn may close the bandgap between upper and lower Jahn-Teller band. The resulting redistribution of electrons from the lower to the upper Jahn-Teller band could eventually lead to a collapse of the orbital polarization. Loss of orbital polarization lowers the Jahn-Teller splitting, and this eventually causes the metal-insulator transition.

Hence, we tend to attribute the orthorhombic-to-pseudocubic transition at $T_{JT}$ to the loss of orbital polarization rather than that of orbital order. With this reassignment, there is no contradiction with an orbital order transition near room-temperature consistent with our findings.

Both transitions, the O/O' and the O/C transition, have a strong tilt component as indicated by the strong dependence of the transition temperatures on the tolerance factor [68]. The thermodynamics of the octahedral tilt pattern, which strongly determines the ordering at high temperatures, is expected to be similar for PrMnO$_3$ and CaMnO$_3$. The latter exhibits two tilt transitions close to 900 °C [70]. It is conceivable, although not guaranteed, that the phase transitions $T^*$ and $T_{JT}$ are connected to the two tilt transitions observed for CaMnO$_3$.

---

[2] In our simulations this broadening happens at discrete points due to the finite super-cell size

[3] Sanchez et al. emphasize that the pseudo-cubic phase is still actually orthorhombic and names it O* rather than C. The assignment of the O/O' transition has been done by us on the basis of the similar data of Sanchez and Pollert.

A direct experimental probe of the orbital order is the resonant x-ray scattering experiments by Murakami et al. [19] on LaMnO$_3$. The orbital order reflection (3,0,0) exhibits a clear drop near 200 K, which the authors relate to the Neel transition at 140 K. Our finding offers the orbital order transition as an alternate explanation for the drop in the orbital order reflection. However, the work by Murakami et al. [19] also shows a high-temperature tail of the reflection. A possible explanation for this tail is a remaining orbital polarization induced by the tilt pattern, which persists up to the orthorhombic to orthorhombic transition of LaMnO$_3$ at 780 K. Furthermore, Zimmermann et al. [17] observed an orbital order (0,3,0) reflection at the Mn K edge in polarized x-ray for Pr$_{0.75}$Ca$_{0.25}$MnO$_3$ which exhibits a step-like reduction above 220 K and a high temperature tail up to 850 K, where it vanishes.

This apparent contradiction can be resolved within a model in the spirit of the Landau theory of phase transitions for two coupled order parameters. As reviewed by Cowley [31], there are several types of Landau free-energies that describe the impact of a primary order parameter, e.g., due to a structural phase transition, on a secondary order parameter, such as an induced physical property. Here, we start from the free energy of two independent second-order phase transitions at different temperatures, one for the orbital order parameter and another one for the tilt angle. The saturation of the order parameters further away from the transition temperature is accounted for by replacing the linear prefactor $\frac{T-T_C}{\Delta}$ by $\tanh\left(\frac{T-T_C}{\Delta}\right)$. Finally, the two order parameters $x$ and $y$ are coupled by a bilinear term.

Let us consider a free energy of the form

(8) $F_T(x,y) = \frac{1}{4}x^4 + \tanh\left(\frac{T-T_x}{\Delta_x}\right)\frac{1}{2}x^2 + \frac{1}{4}y^4 + \tanh\left(\frac{T-T_y}{\Delta_y}\right)\frac{1}{2}y^2 - \alpha xy$

which describes the free energy for two order parameters x and y as function of temperature $T$. For $\alpha$=0, the free energy describes two systems that each undergo a second order phase transition at, respectively $T_y$. The parameters $\Delta_x$ and $\Delta_y$ control the sharpness of the corresponding transition. The last term with the coupling parameter $\alpha$ describes the coupling of the two systems. For $\alpha$>0 it favors states with order parameters having equal sign over those with opposite sign. The parameters ($T_x$=300 K, $\Delta_x$=50 K, $T_y$=700 K, $\Delta_y$=100 K and $\alpha$=0.9 have been adjusted so that the orbital-order parameter resembles the data for the orbital reflection of Murakami et al. [19] The variable $x$ represents the order parameter for the orbital order, while the variable $y$ represents the tilt angle. For the sake of simplicity, we consider only one tilt angle. Both order parameters are given in units of their uncoupled low-temperature values.

As shown in Fig.11, in the presence of a coupling, the order parameter $x$ for orbital order has two non-zero plateaus. At low temperatures the orbital order is spontaneous, while being enhanced in size due to the cooperative effect of the tilt angle. At the temperature of the orbital order transition of the uncoupled system, the order parameter does not fall off to zero, but it exhibits a drop to a lower value, which is due to orbital order induced by the tilt pattern. Consequently, $T_x$=300 K marks the temperature, for which the orbital order is largely lost, i.e. the transition temperature for spontaneous orbital ordering. While the order parameter x for the orbital order goes to zero at $T_y$=800 K simultaneously with that for the tilt pattern, one would characterize the region of the induced orbital order by the tilt pattern rather than by the orbital order.

Interesting are the changes upon doping, which reduces the weight of the checkerboard pattern of the orbital order by inserting hole polarons. Hole polarons in the background of the checkerboard orbital order is shown in Fig.1. This reduces the magnitude of the induced orbital order, as opposed to shifting the transition temperature from $T_y$=800 K to $T_x$=300 K , as one might expect.[4]

---

[4] The hole polarons have a destabilizing effect, which lowers the orbital-order transition temperature, but this effect, which is also present in our simulations, shifts the transition temperature by a much smaller amount and it is independent of the tilt pattern.

When, the hole polarons order themselves in a pattern that cannot couple to the tilt pattern, the two transitions at $T_x$ and $T_y$ become decoupled and two independent second-order transitions is obtained. This is clearly seen by comparing the temperature dependence of Zimmermann et al. [17]: The orbital (0,3,0) reflection for doping PCMO x=0.25 exhibits an induced orbital order, while that for doping PCMO x=0.4 in the charge-ordered regime undergoes an abrupt transition to zero at 240 K.

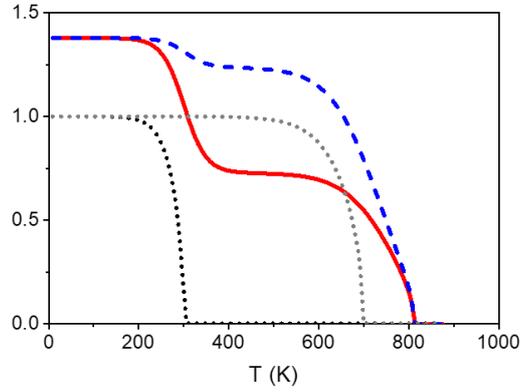

**FIG. 11**. Order parameters for the free energy of Eq.8 as function of temperature. The full line describes the orbital order parameter *x*, and the dashed line represents the tilt angle. Both are scaled to their low-temperature value without coupling. The dotted lines show the order parameters in the absence of coupling.

The distinction between induced and spontaneous order in the experimental studies has an impact on the classification of the phase transition. Fig. 9 shows continuous changes of the lattice parameter across the phase transition, indicative for a continuous, e.g. second order phase transition. This is also reflected in the fit of the lattice parameter by the structure model including tilt and Jahn-Teller distortion. Only the amplitudes of the $Q_2$ and $Q_3$ modes are changing across the phase transition, whereas the symmetry of the distortions is the same above and below the phase transition. As a result, the lattice possesses *Pbnm* symmetry above and below the orbital order transition. In contrast, the high temperature tilt phase transition is of first order, since both the lattice symmetry is changing as well as the lattice parameters and the cell volume shows a step at the transition temperature.

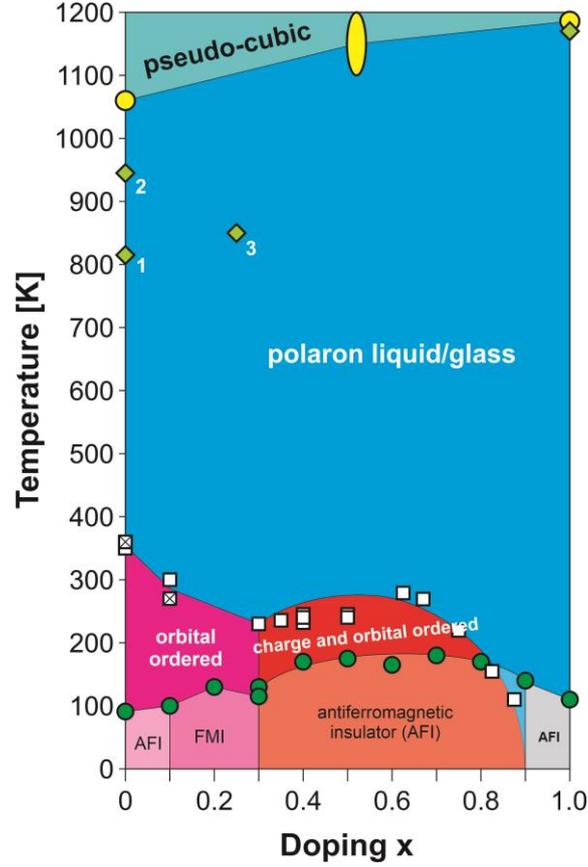

**FIG. 12:** Proposed changes to the phase diagram of PCMO: Data points included in the phase diagram are based on the following published results [7,8,12,16,17,70–73] and listed in more detail in Appendix C. For undoped x=0 and low doped x=0.1 our experimental results based on temperature anomaly of lattice parameters are shown (squares). In addition, the results from theoretical simulations (crossed out squares) redefine the low doped region. Additional data points for the high temperature region have been included: For undoped PCMO x=0 (1) observed change in thermoelectric power [30] as well as (2) a O/O' transition [68] and for PCMO x=0.25 (3) residual orbital ordering measured by resonant X-ray spectroscopy [74]. Note that various magnetic sub phases see refs. [8,12,37] are not indicated. In addition, changes in properties of the polaron liquid phase as a function of doping are not indicated such as possible polaron glass found in other manganites [75] induced orbital order discussed in the text or phase separation close to the charge/orbital order transition [24].

## IV. Conclusions

A previously unknown phase transition has been detected in the low doping region of $Pr_{1-x}Ca_xMnO_3$ at $T_{OO}$=350 K for x=0, $T_{OO}$=300 K for x=0.1 and at approximately $T_{OO}$=220-250 K in thin films with x=0.1. The experimental fingerprints are an anomalous thermal expansion behavior in the bulk materials and subtle, sometimes impactful changes in the physical properties of the thin films. Remarkable is the increase of lifetime of polaron excitations below $T_{OO}$ that corresponds to an onset of the polaron photovoltaic effect.

Our theoretical simulations provide evidence that this transition is the orbital-order phase transition with the appearance of spontaneous orbital ordering. In the theoretical simulations, the orbital order vanishes during the transition while the orbital polarization remains intact up to high temperatures. Our findings indicate that the orbital-order transition in the low-doped region occurs in a similar temperature range for the whole doping regime.

A remaining induced orbital order above the orbital-ordering temperature observed in resonant x-ray diffraction [19] for $LaMnO_3$ and PCMO for x=0.25 [17] is attributed to an induced orbital order caused by octahedral tilt pattern. The tilt pattern couples to the Jahn-Teller distortion and thus induces an orbital order via displacements of the A-type (Pr, Ca) ions. The resonant x-ray scattering studies at PCMO x=0.25 exhibit a strong decay of the orbital-order reflection above 220 K with a tail extending to high-temperatures of 850 K.

Our reassignment of the orbital order transition to appear at temperatures close or below room temperature has an impact on the interpretation of the high temperature phase transition at 950 K, previously assigned to the orbital order transition [13] and taken up by [17,74]. We attribute it to a combination of a tilt transition with a metal insulator transition, which explains the increased metallic, conductivity above the phase transition. The altered phase diagram is show in Fig. 12.

Other studies of orbital order in the charge- and orbital ordered range of the phase diagram of PCMO using resonant x-ray diffraction never found an orbital ordered state at temperature close to the melting temperature and thus support our conclusion that spontaneous orbital ordering driven by electronic correlations is only emerging at low and moderate temperatures. Our results suggest also to reevaluate the orbital ordered corner of the phase diagrams of other orthorhombic manganites such as $LaMnO_3$ and $NdMnO_3$.

**Acknowledgement:**

This work is funded by the Deutsche Forschungsgemeinschaft (DFG, German Research Foundation) - 217133147/SFB 1073, projects B02, B03, A06 and B07. The authors thanks K. Stroh for supporting the absorption experiments. MTB and PB are grateful to Dr. Sangeeta Rajpurohit for fruitful discussions on the dynamical and thermodynamical properties of manganites.

B. Kressdorf, T. Meyer and M. ten Brink contributed equally to this work.

# APPENDIX A – Experiments

**Sputter deposition of epitaxial films:**

Epitaxial $Pr_{0.9}Ca_{0.1}MnO_3$ (PCMO) thin films were deposited on (100) $SrTiO_3$ (STO) and (100) $SrTi_{0.995}Nb_{0.005}O_3$ (STNO) substrates by ion-beam sputtering with a thickness of 100 nm. STNO serves as the n-doped part of the heterojunctions. For optical and pump-probe spectroscopy epitaxial PCMO films on MgO substrates were used, because of the larger bandgap and higher thermal conductivity of this substrate.

During deposition the total gas pressure amounted to $5.4 \times 10^{-4}$ mbar. Argon was used for beam neutralization ($P_{Ar}=3 \times 10^{-4}$ mbar) and Xenon for sputtering ($P_{Xe}=1 \times 10^{-4}$ mbar). An oxygen partial pressure of $P_{O2}=1.4 \times 10^{-4}$ mbar is sufficient for a correct oxygen stoichiometry of the perovskite phase. The deposition temperature of the substrate surface was approximately 720° C (the temperature of the heater was fixed at 820° C). In order to reduce strain and growth-induced defects that affect resistivity, the samples for electric transport measurements were post-annealed in air at 900° C up to 20 h. The heating and cooling rates were fixed to 100K/hour.

**X-ray diffraction (XRD):**

In order to characterize the out-of-plane growth directions and the strain states of the thin film samples, XRD in $\Theta-2\Theta$ Bragg-Brentano geometry and at room temperature was performed with a Bruker D8 Discover system with monochromatic Cu $K\alpha_1$ radiation $\lambda=1.5406$ Å.

For temperature-dependent characterization in the range between 540 K and 90 K, polycrystalline bulk powders prepared by conventional solid-state reaction were investigated using a Bruker D8 Advance Powder X-Ray system with Cu $K\alpha$ radiation. In order to avoid oxygen losses, measurements above room temperature were performed in air. For low-temperature measurements the sample holder was cooled with liquid nitrogen and the sample chamber was evacuated down to $P=10^{-2}$ mbar. The lattice parameters were deduced from measurements with an angular resolution of 0.01°.

**Photovoltaic characterization:**

For photovoltaic characterization, previously reported sample geometry was used [10]. On the rear side of the STNO substrate, ohmic back contacts were realized by Ti layers with protective Au coatings. Using a shadow mask, Pt top contacts with a thickness of 200 nm and a size of 4 mm x 1 mm were deposited onto the PCMO x=0.1 films. All current densities are normalized to this top contact area. The deposition of the contact layers was also performed by IBS.

The samples were set up in a Cryostat (Cryovac) with a suprasil entry window and analyzed in a temperature range from 80 K to 300 K. Excluding resistance contributions of the supply cables, top and back contacts were connected to a Keithley 2430 serving as voltage source and ammeter. Additionally, the voltage drop was measured by a Keithley 2182A Nanovoltmeter (internal resistance > 10G Ω).

A LOT 150W Xe-UV lamp served for polychromatic illumination and, by introducing cut-off filters characterized by a maximum photon excitation energy $E_{max}$, for illuminations with $E_{ph} \leq E_{max}$. The output power of the source was set to 175 mW.

**Temperature-dependent resistivity:**

Temperature-dependent four-point resistance measurements in zero magnetic field and at 9 T at a constant voltage of 75 mV were performed in a Physical Properties Measurement System (PPMS) from Quantum Design. The investigated PCMO x=0.1 films on insulating STO-substrates were prepared with ion-beam sputtered 4x1 mm Pt top contacts with a spacing of 2 mm by using a shadow mask.

**Magnetic properties**:

The field- and temperature dependent magnetic moment of PCMO x=0.1 samples on STO has been measured with a Quantum design SQUID-magnetometer. For all measurements, the DC-measurement mode had been used. Diamagnetic background contributions of the sample holder and the STO substrate were deduced from a blank (100) STO substrate. This background signal was then subtracted from the experimentally observed moment of coated substrates.

**Optical spectroscopy:**

The overview of the spectral- and temperature-dependent absorption coefficient was measured in a transmission setup and is not corrected by reflection. In the UV-Vis range ($E_{ph}$=1.13 - 6.2 eV) the setup consists of an "OceanOptics DH-2000" Halogen light source, a fiberglass QP400-2-SR-BX and a Maya2000Pro spectrometer. For the determination of the band gab additional spectral- and temperature-dependent reflection measurements in the range $E_{ph}$=1.2–1.6 eV were included to calculate a reflection corrected absorption coefficient. For the reflectance measurements enhanced AU mirror (Thorlabs) is included in the setup, which provides a reference spectrum.

The transmittance was calculated by $I=I_T/I_0$, where $I_0$ is the incident and $I_T$ is the transmitted spectrally resolved intensity. A baseline correction was applied to the detectors by subtracting a dark spectrum.

**"Pump-probe" experiments**:

Experimental data on transient transmission was recorded with a bichromatic pump-probe setup making use of a femtosecond fiber amplifier system from Active Fiber Systems running at 50 kHz repetition rate. The pump beam has a central wavelength of 1030 nm (1.2 eV photon energy), the frequency-doubled probe beam has a central wavelength of 515 nm (2.4 eV photon energy). Both pump and probe beam have a pulse duration of less than 40 fs, as checked by autocorrelation in front of the sample. The measurements are performed in nearly collinear beam geometry inside a liquid helium flow cryostat from Janis with optical access. In order to obtain high signal to noise ratio, the pump beam is additionally modulated with an external chopper at a frequency of 534 Hz to allow direct detection of the transmission change $\Delta Tr$ using lock-in detection at the chopper frequency. Probe signal detection itself is performed using a silicon photodiode with an added bandpass filter to block residual pump light. The measurements were performed on an PCMO x=0.1 sample on MgO [76].

**APPENDIX B: Details of the simulation:**

**Model parameters**

The parameters for the model Hamiltonian are extracted from density-functional calculations of $Pr_xCa_{1-x}MnO_3$ with $x \in \{0, \frac{1}{2}, 1\}$. The energy terms and the parameters of the model Hamiltonian are provided in an earlier publication [46].

We modified this Hamiltonian in the following ways: (1) We allow the lattice constants to adjust dynamically, as mentioned above, but we constrain the lattice constants in the *ab*-plane to be equal. (2) Following Rajpurohit et al. [77] the resting force constant of the breathing mode is reduced slightly from $k_{br} = 10.346$ eV/ Å$^2$ to 9.04 eV/ Å$^2$ to reproduce the ratio of the amplitudes of Jahn-Teller vs. breathing mode of the ab-initio calculations for PrMnO$_3$.

**Computational details**

The Car-Parrinello dynamics [58] used in our simulations introduces fictitious kinetic energy terms for all dynamic degrees of freedom. The degrees of freedom are described by wave functions $\Psi_{\sigma,j,R,n}$ for the e$_g$ electrons, occupations $f_n$, positions $R_{R,R'}$ connecting manganese site $R$ to the neighbor $R'$, spins $\vec{S}_R$ and lattice constants $T_{x/y/z}$. Here $n$ refers to the band index, index $R$ to manganese sites, σ and $j$ to spin and orbital degrees of freedom of the wave functions. This yields the Lagrangian:

$$\mathcal{L} = \sum_{\sigma,j,R,n} f_n \langle \dot{\Psi}_{\sigma,j,R,n} | m_\Psi | \dot{\Psi}_{\sigma,j,R,n} \rangle + \frac{n}{2} \sum_{R,R'} M_R \dot{R}_{R,R'}^2$$
$$+ \frac{1}{2} \sum_R m_S \dot{\vec{S}}_R^2 + \frac{1}{2} \sum_{k \in \{x,y,z\}} M_T \dot{T}_k^2 - E_{pot}[\Psi, f, R, S, T] - \mathcal{L}_{constr.}$$

Here the dot refers to the time derivative, $E_{pot}$ summarizes all energy terms introduced before and described in Ref. [46] and the last term $\mathcal{L}_{constr.}$ describes additional constraints, such as orthogonality of the wave functions. Friction terms have been added to the Euler-Lagrange equations.

In the finite-temperature calculation, the Nosé-Hoover thermostat [59,60] drives the thermal fluctuations via a (positive and negative) friction acting on the oxygen atoms. The electrons and spins follow the atoms quasi-adiabatically, when they have a sufficiently small mass and a small constant friction. The values of the thermostat masses, frictions and the period of the Nosé-Hoover thermostat are given in table 1. The lattice constants are kept fixed during the finite temperature calculations. For the ground state calculation, the masses and frictions are adjusted for a rapid convergence and the lattice constants are optimized.

The finite temperature calculations are performed in a 4x4x4 manganese-sites unit cell. We use a doping of $x = \frac{6}{4^3} = 0.09375$, close to the nominal experimental doping of x=0.1. Calculations in a 6x6x4 sites unit cell at doping $x = \frac{14}{144}$ confirmed the ground state and the transition temperature obtained in the smaller unit cell.

| $m_\Psi$ | $m_S$ | $M_R$ | $P_{TS}$ | $f_\Psi$ | $f_S$ |
|---|---|---|---|---|---|
| $1.0\, m_e a_0^2$ | $1.0\, m_e \cdot \left(\frac{a_0}{3/2\hbar}\right)^2$ | 15.999 u | 100 fs | $0.41\, \frac{1}{fs}$ | $0.41\, \frac{1}{fs}$ |

Table 1: Masses and friction values used in the Lagrangian Equation. $m_\Psi$ and $m_S$ are the fictitious masses of electron wave functions and spins. $M_R$ is the oxygen mass. $P_{TS}$ is the quasi-period of the Nosé-Hoover thermostat. The friction forces are $-f_\Psi m_\Psi \dot{\Psi}_{\sigma,\alpha,R,n}$ for the wave functions and $-f_S m_S \dot{S}_R$ for the spins.

**APPENDIX C: Phase diagram for PCMO**:
Data points included in the phase diagram in Fig. 10 are based on data in this publication and the following previously published results:

**Low temperature magnetic transitions**

Summary of the magnetic transitions documented by Jirak et al. for doping levels PCMO x=0-0.9 and Wollan and Koehler for PCMO x=1 in a combination of x-ray and neutron diffraction measurements as well as additional transport measurements [8,12,16]:

- x=0: $T_N$=91 K antiferromagnetic type A order.
- x=0.1: $T_N$=100 K ferromagnetic order with easy axis [010], below $T_2$=70 K canted antoferromagnetic order.
- x=0.2: $T_C$=130 K ferromagnetic order.
- x=0.3: ferromagnetic and antiferromagnetic contributions $T_C$=115 K and $T_N$=130 K respectively.
- x=0.4: $T_N$=170 K antiferromagnetic CE type.

- x=0.5: $T_N$=175 antiferromagnetic CE type.
- x=0.6: $T_N$=165 K antiferromagnetic CE type.
- x=0.7: $T_N$=180 K antiferromagnetic and noncommensurate spiral magnetic ordering.
- x=0.8: $T_N$=170K antiferromagnetic type C.
- x=0.9: $T_N$= 140 K antiferromagnetic type C and T=110 K antiferromagnetic type G.
- x=1.0: $T_N$=110 K antiferromagnetic type G.

**Charge and orbital order transitions at medium temperature:**

Our bulk analysis of the lattice constant determined by temperature dependent x-ray diffraction suggests an orbital-order transition temperature of 350 K and 300 K for PCMO doping levels of x=0 and x=0.1, respectively. For higher doping levels between PCMO x=0.3 and PCMO x=0.5, a simultaneous transition of orbital and charge ordering has been experimentally observed in a variety of the following publications and measuring techniques resulting in similar transition temperature values. For example, Zimmermann 2001 et al. preformed resonant x-ray spectroscopy measurements to determine a simultaneous charge and orbital ordering transition at $T_{CO/OO}$=245 K for doping of PCMO x=0.4 and PCMO x=0.5 [17]. The structural analysis and transport measurements performed by Jirak et al. [8,12] determines the charge order / orbital order transition at $T_{CO/OO}$=230 K for PCMO x=0.3 and $T_{CO/OO}$=232 K for PCMO x=0.4, whereas Yoshizawa et al. [71] gives a transition temperature of $T_{CO/OO}$ = 200 K based on change in lattice parameter determined from neutron diffraction studies for PCMO x=0.3. In temperature dependent resistivity measurements, the onset of charge order is visible in a pronounced step in resistivity. Based on such measurements, Tomioka et al. [7] determined $T_{CO/OO}$ =220-230 K for PCMO x=0.35/0.4/0.5 and T=200 K for PCMO x=0.3

The charge order materials are generally considered to show a broad two phase region of order and disorder domains. Jooss et al. and Wu et al. showed for PCMO x=0.32 and x=0.5 a two phase region of charge order / orbital order domains visible from room temperature until 70 K [24,27].

For higher doped region PCMO x=0.5 to x=0.875 Zheng et al. measured transport, ultrasound and powder x-ray diffraction [73]. In this study samples with the following doping have been investigated PCMO x=0.5, 0.55, 0.6, 0.625, 0.67, 0.7, 0.75, 0.8, 0.825, 0.85, 0.875.

**Structural phase transitions at high temperature:**

As previously mentioned for PCMO x=0, a variety of publications on the high-temperature phase transition exist. The exact transition temperatures and the statement whether there are one or two transitions varies. Pollert et al. published neutron diffraction measurements indicating a O to O' to pseudocubic transition for $T_{OO'}$=815 K and $T_{O'C}$=945 K, respectively [13]. Whereas Sanchez published considerable higher $T_{JT}$ at 1050 K and a possible second transition at 948 K [30], Zhou and Goodenough identified two phase transitions using a combination of resistivity and thermoelectric power measurements. They yield $T_{JT}$ and $T^*$ at around 1050K and 750 K [68]. Even though the absolute temperatures vary, we suggest that all three publications observe the same two transitions although their assignment differs.

For doping PCMO x=0.52 Carpenter et al. published a symmetry and strain analysis of structural phase transitions using powder neutron diffraction [72]. At high temperatures a first order transition of orthorombic *Pbmn* structure to $R\bar{3}c$ is reported in a temperature region between 1100 and 1200 K, which has a significant hysteresis [72]. Zimmermann et al. analyses resonant x-ray scattering for PCMO x=0.25 and finds that the orbital ordered (030) peak has a residual intensity until 850 K [17]. In case of PCMO x=1.0 Taguchi et al. [70] found a orthorhombic phase with a transition into tetragonal phase at 1170 K and a subsequent cubic transition at 1186 K.

# Orbital order phase transition in Pr$_{1-x}$Ca$_x$MnO$_3$ probed by photovoltaics

B. Kressdorf[1], T. Meyer[2], M. ten Brink[4], C. Seick[3], S. Melles[1], N. Ottinger[1], T. Titze[3], H. Meer[3], A. Weisser[3], J. Hoffmann[1], S. Mathias[3], H. Ulrichs[3], D. Steil[3], M. Seibt[2], P.E. Blöchl[5], C. Jooss[1]

[1] *University of Göttingen, Institute of Materials Physics,*
[2] *University of Göttingen, 4th Institute of Physics*
[3] *University of Göttingen, 1st Institute of Physics*
[4] *University of Göttingen, Institute of Theoretical Physics*
[5] *Clausthal University of Technology, Institute of Theoretical Physics*

## I. Microstructural characterization of epitaxial thin film junctions

Fig. S1 shows the structural characterization of a typical Pr$_{0.9}$Ca$_{0.1}$MnO$_3$ (PCMO x=0.1) -STNO heterojunction by x-ray diffraction (XRD) measurements and transmission electron (TEM) studies. In the Pbnm notation, epitaxial relations typically observed for PCMO films on (001) STO are

[001]$_{oop}$ STO || [001]$_{oop}$ PCMO with [110]$_{ip}$ STO || [100]$_{ip}$ PCMO or [110]$_{ip}$ STO || [010]$_{ip}$ PCMO

and

[001]$_{oop}$ STO || [110]$_{oop}$ PCMO with [100]$_{ip}$ STO || [001]$_{ip}$ PCMO or [010]$_{ip}$ STO || [001]$_{ip}$ PCMO

Here, oop corresponds to the epitaxial relation perpendicular to the substrate and ip to the in-plane configuration. Additional exchange of the *a* and *b* directions of PCMO leads to six different twin domains.

The XRD scan in Fig. S1 (a) reveals that PCMO x=0.1 has two out-of-plane growth directions, [220] and [004], that corresponds to the usually observed twinning. No additional orientations are visible. By considering the different structure factors of [220] and [004], the intensity ratio of the experimentally observed peaks implies predominantly [001] growth with a volume fraction of about 85%. This has been confirmed by TEM annular dark-field (ADF) imaging of cross plane lamellas.

Electron-transparent lamellae in cross-section geometry were prepared by means of focused ion-beam etching (FEI Nova NanoLab Dual Beam system) using an acceleration voltage of 5 kV during the final thinning step. The TEM measurements were carried out in an FEI Titan 80-300 operated at 300kV and equipped with a Gatan Quantum 965 ER image filter. Selective area electron diffraction (SAED) was performed with a 10 μm aperture that approximately corresponds to a lateral size of about 170 nm in the image plane. Scanning TEM (STEM) images with an annular dark field (ADF) detector were taken at an electron current of 42 pA with an inner and outer acceptance semi-angle of 46.8 mrad and 200 mrad, respectively (camera length 38 mm). Electron energy loss spectroscopy (EELS) data acquisition was performed at a beam current of 150 pA and at a collection semi-angle of 39 mrad.

Fig. S1 (b) shows a typical columnar pattern of (001) oriented quite regularly arranged domains with a size of 100-150 nm. Remarkably, the interface between STO and PCMO x=0.1 is almost coherent with a very low density of dislocations (Fig. S1 (c)), although the nominal lattice mismatch is rather large. The formation of [220] and [004] orientational twin domains is a result of the local symmetry breaking in the cubic to orthorhombic phase transition above 600 K [1]. In addition to the (minor) contribution of (220) orientations they are visible by subsequent alternation of PCMO x=0.1 in-plane [100] and [010] orientations within the (001) oriented domains. Post-annealing of the sample at 900° C gives rise to a quite strong relaxation of the small volume fraction of (220) oriented domains and a pronounced reduction of the out-of-plane strain ε$_{oop}$ (Fig. S1 (a)).

The strain stress state in our thin films can be determined through the XRD measurement. According to [2], the lattice parameters of PCMO x=0.1 are *a*=0.5442 nm, *b*=0.5617 nm and *c*=0.7635 nm. Using the lattice constant of STO of 0.3905 nm, the lattice mismatches for [001] growth correspond to $\varepsilon_{ip}(a)$ = 1.48 % ($[110]_{ip}$ STO || $[100]_{ip}$ PCMO) and $\varepsilon_{ip}(b)$ = -1.68 % ($[110]_{ip}$ STO || $[010]_{ip}$ PCMO).

From simple isotropic elastic consideration, the out-of-plane strain $\varepsilon_{oop}$ should be of the order of

$$\varepsilon_{oop} \approx -\frac{\nu}{1-\nu}\left(\varepsilon_{ip}(a) + \varepsilon_{ip}(b)\right), \quad (1)$$

where Poisson's ratio of PCMO is about 0.3. Therefore, one would expect an out-of-plane strain of the order of 0.1% which is considerably smaller than the experimentally observed $\varepsilon_{oop}$ = 0.8% deduced from the plane spacing of the [004] reflection. This suggests an additional compressive in-plane strain component. Since this additional component almost vanished after post-annealing (see main text), preparation-induced defects are the most likely source of this strain.

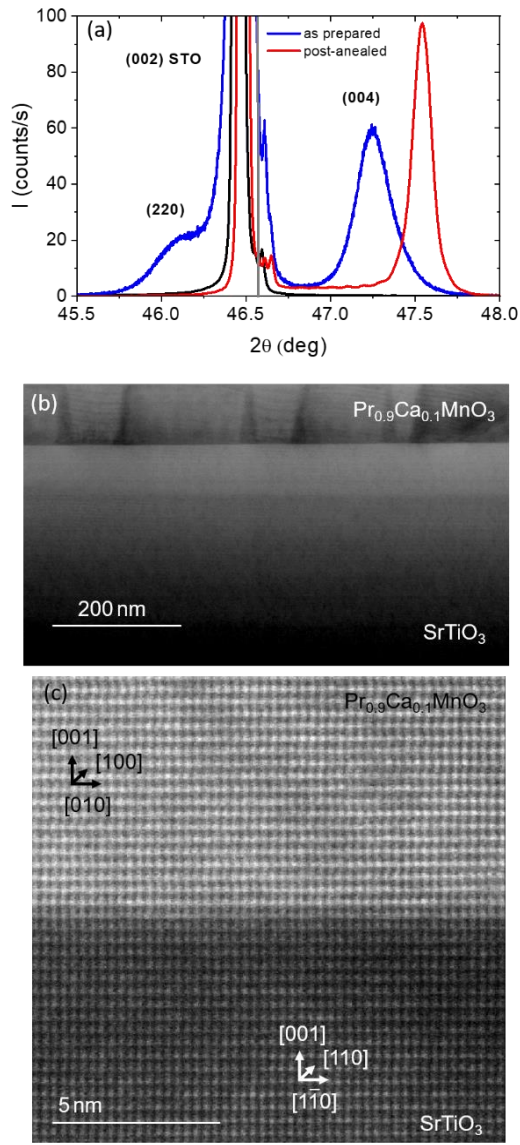

**FIG. S1:** Structural characterization of the PCMO x=0.1 / STO heterojunction. (a) XRD of the PCMO x=0.1 film on (001) STO, as-prepared (blue) and post-annealed for 4h at 900° C in air. The black lines indicate the angular peak positions of (220) and (004) for bulk PCMO x=0.1 [2]. (b) ADF-STEM image showing a nano twinned PCMO x=0.1 films on STO. The nano-beam electron diffraction pattern of the red and blue marked areas show that all visible twins have (001) orientation with respect to the substrate and changing [010] resp. [100] direction

parallel to the STO [110] zone axis (see SI). (c) ADF-HRSTEM of a [100] twin domain showing a coherent interface between film and substrate.

## II. Resistivity and logarithmic derivative

Post-annealing, i.e. annihilation of preparation-caused defects, strongly affects the temperature dependent resistivity of weakly doped PCMO x=0.1. Fig. S2 (a) shows the changes of resistivity $\rho$ due to post-annealing for $t_a$ hours at 900° C. The heat treatment causes a reduction of resistivity by roughly one order of magnitude. The magneto-resistance effect

(2a) $\quad \frac{\Delta\rho}{\rho} \equiv \frac{\rho(0)-\rho(9\,T)}{\rho(0)}$

at room temperature amounts to about 6% but strongly increases to about 30% at 180 K. The effect of post-annealing is small but an increase with increasing annealing time is observed.

Fig. S2 (b) shows the influence of different annealing times on the logarithmic derivative

(2b) $\quad E_A = kT \frac{d}{dt} \ln\left(\frac{\rho}{T}\right)$

that equals the small polaron hopping energy in the adiabatic approach.

As mentioned in the main text, the main features, such as the nearly linear decrease of E with decreasing temperature, as well as the reduction of the slope due to post-annealing, the appearance of a peak and the suppression of this peak in strong magnetic fields, are also observed experimentally in PCMO x=0.34 thin films, which, like PCMO x=0.1 studied here, undergo a low-temperature transition to a charge and orbital ordered state. In [3], the interplay of preparation-caused defects, charge- and orbital ordering and melting in strong magnetic fields is discussed.

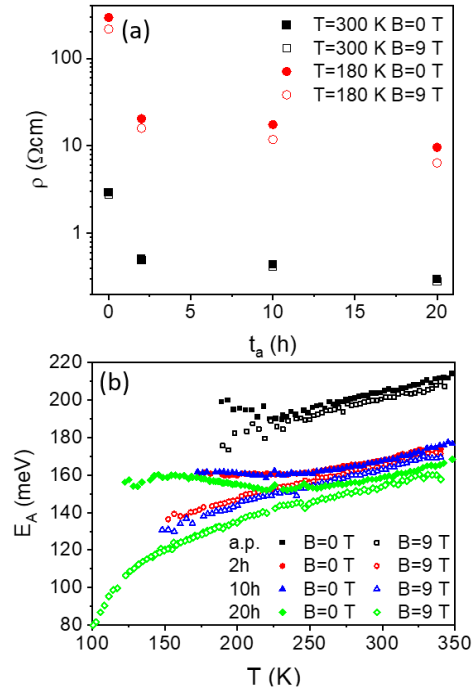

**FIG. S2:** (a) Resistivity at 300/180 K in zero field and $B$=9 T vs. annealing time $t_a$. Post-annealing was performed at 900° C with a heating and cooling rate of 100 K/min. (b) Logarithmic derivative $E_A$ vs. temperature in zero field and $B$=9 T after different annealing time $t_a$.

## III. Reflection correction of the absorption coefficient

Using Taucs's relationship, the reflection-corrected absorption coefficient $\alpha$ and the optical band gap $E_g$ can be deduced from the transmittance $T$, the reflectance $R$ and the film thickness $d$ [3 and references within]:

(3a) $\quad \alpha = \frac{1}{d} \ln \left[ \frac{(1-R)^2}{2T} + \sqrt{\frac{(1-R)^4}{4T^2} + R^2} \right]$

(3b) $\quad \alpha h\nu = \alpha_0 (h\nu - E_g)^n$

$\alpha_0$ is the band tailing parameter and $n$ is a characteristic exponent depending on the character of transition. Fig. S3 (a) shows the absorption coefficient according to equation (3a) for various temperatures in the photon energy regime of $E_{ph}$=1.2 eV to $E_{ph}$=1.6 eV. The reflection data for the samples were measured and scaled down by 7% to match a reference mirror sample. In order to determine the band gap, we compare the measured $\alpha h\nu$ vs $h\nu$ with Eq.3b for various trial values of the band gap $E_g$ and for the characteristic exponents of $n$= 1/2 for direct allowed transitions, $n$=2 for indirect allowed, $n$=3/2 for direct forbidden and $n$=3 for indirect forbidden transitions. The exponent $n$=3/2 gave the best fit.

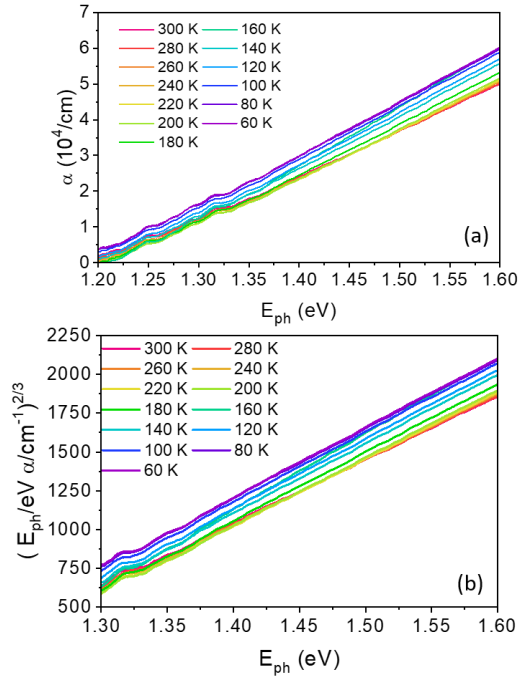

**FIG. S3**: (a) Temperature and spectral dependence of the absorption coefficient deduced from eq. 3a. (b) Linearized plot according to eq. 3b with characteristic exponent $n$=3/2.

### IV. Magnetic measurements: Substrate correction

As mentioned in the main text, the paramagnetic moment of PCMO x=0.1 films on (100) SrTiO$_3$ (STO) can only be deduced from SQUID measurements if the substrate contributions are considered.

As anticipated, the blank substrate measurement $m_{STO}$ exhibits a mainly diamagnetic contribution. However, it weakly depends on temperature and applied field, indicating additional magnetic contributions, that are, most probably, due to magnetic impurities. We have measured the magnetic moment of the substrate before deposition and of the stack of film with the identical substrate after deposition. Since the ferromagnetic surface contribution of STO substrates may change while heating to the PCMO x=0.1 deposition temperature, the paramagnetic moment $m_{film}$ of PCMO x=0.1 cannot directly be related to the difference between the total moment $m_{tot}$ of film and substrate and the substrate moment $m_{STO}$.

Therefore, the difference was modelled according to

(4) $\quad m_{tot}(T) - m_{STO}(T) = \Delta m_s + m_p(T) = \Delta m_s + \frac{\chi_0 H}{T - T_{CW}}$

As shown in the main text, the analysis yields to a paramagnetic PCMO susceptibility that indeed reveals Curie-Weiss behavior with small differences between the orbital ordered and disordered phases. Fig. S4 summarizes the different contributions to the magnetic moment at an external field of 200 mT. Since all contributions are comparable in magnitude the extraction of the film contribution needs careful consideration.

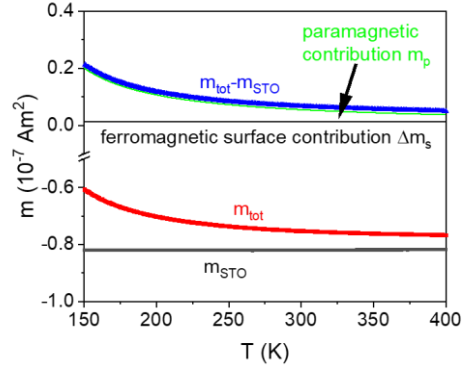

**FIG. S4:** PCMO x=0.1 film on SrTiO$_3$ (STO): Contributions to the magnetic moment at $B_0$=200 mT. $m_{STO}$: blank STO substrate (black symbols). $m_{tot}$: PCMO x=0.1 on STO (red symbols). Difference $m_{tot}$-$m_{sto}$ (blue symbols). $m_p$: paramagnetic PCMO x=0.1 contribution (green line). $\Delta m_s$: temperature-independent contribution due to STO surface ferromagnetism (black line).

## V. Simulation of pump probe dynamics

Finite-Difference Time Domain (FDTD) simulations of a one dimensional three-temperature model [5] were carried out in order to interpret the pump-probe experiments. In this model, the PCMO film (100 nm) on a MgO substrate (1000 nm) is divided into three subsystems, conventionally associated with electrons, lattice and spins. To each subsystem, a temperature is assigned, which evolves in time and space according to a simple diffusion equation. Differential coupling terms $\sim(T_i - T_j)$ assure equilibration of the subsystems.

Note that free electrons are likely not present in the PCMO film. We here include them, representing an energy reservoir which a) takes up the optically deposited energy, and b) rapidly (faster than a few picoseconds) equilibrates with the lattice. These simulations explore a counter argument to the explanations given in the main text, namely, the outcome from our pump-probe experiments on time scales above 10 ps, when only thermal diffusion inside the manganite and into the substrate are considered.

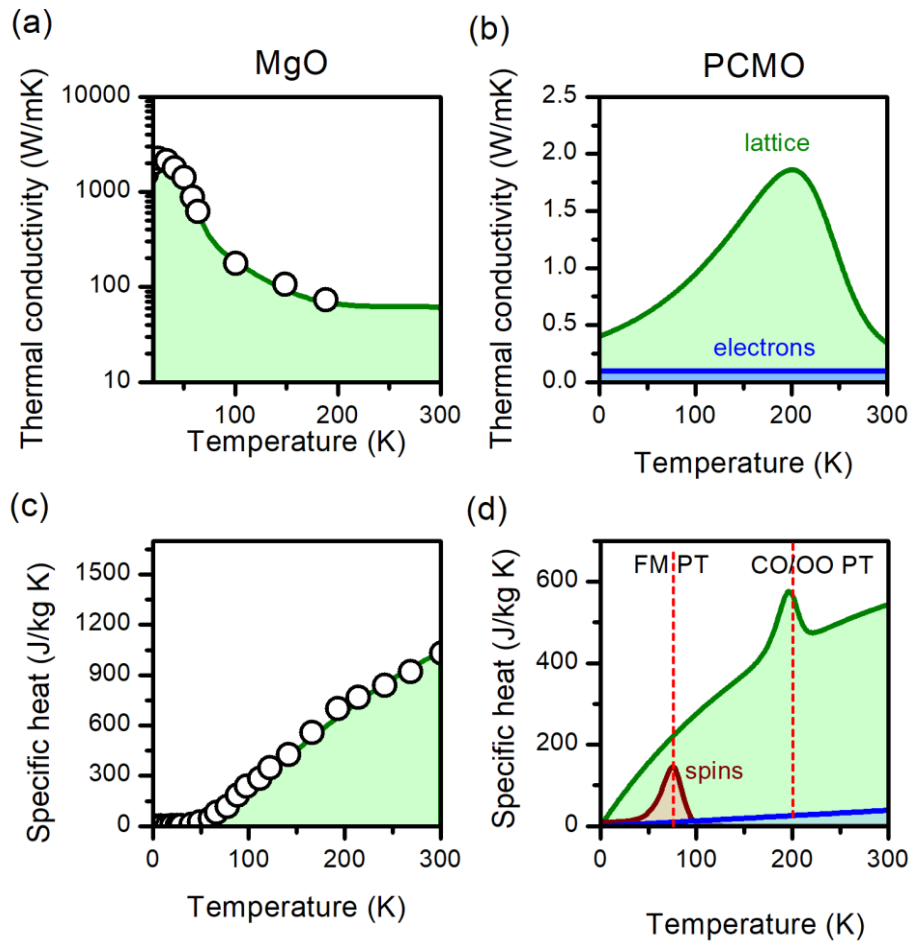

**FIG. S5**: Temperature dependencies of modelling parameters used for the FDTD simulation. (a) and (c) refer to MgO. The green curves interpolated between the experimental data points from […] and […] is used for simulation. For the non-magnetic insulator MgO only the lattice temperature is defined. (b) and (d) refer to PCMO, for which below the FM PT at 80 K three temperature reservoir (electrons, lattice, spin) are distinguished. The CO or OO PT at 200 K is modelled by a corresponding local maximum in the lattice specific heat.

When only thermal diffusion is considered, thermal relaxation is governed by the thermal conductivity and the specific heats of the manganite film and of the MgO substrate, including their temperature dependencies. Figure S5 shows these temperature dependencies entering the coupled diffusion equations. For MgO (see Fig. S5 (a) and (c)), tabulated data from literature were taken and interpolated [6] or the special PCMO x=0.1 composition investigated in this work, such data do not yet exist. Therefore, the generic behavior of the thermal material properties (see Fig. S5 (b) and (d)) compatible with typical experimental findings obtained from different manganite films was assumed. [7,8] During the simulation, the parameters are adjusted in each time step to the local temperature. Note that, our FDTD model [5] was initially developed to describe pump-probe reflectivity experiments. In contrast, the experiments discussed in the main article are concerned with pump-probe transmission. The physical processes sampled by these two measurement schemes are nevertheless almost identical, in particular regarding transient thermal properties.

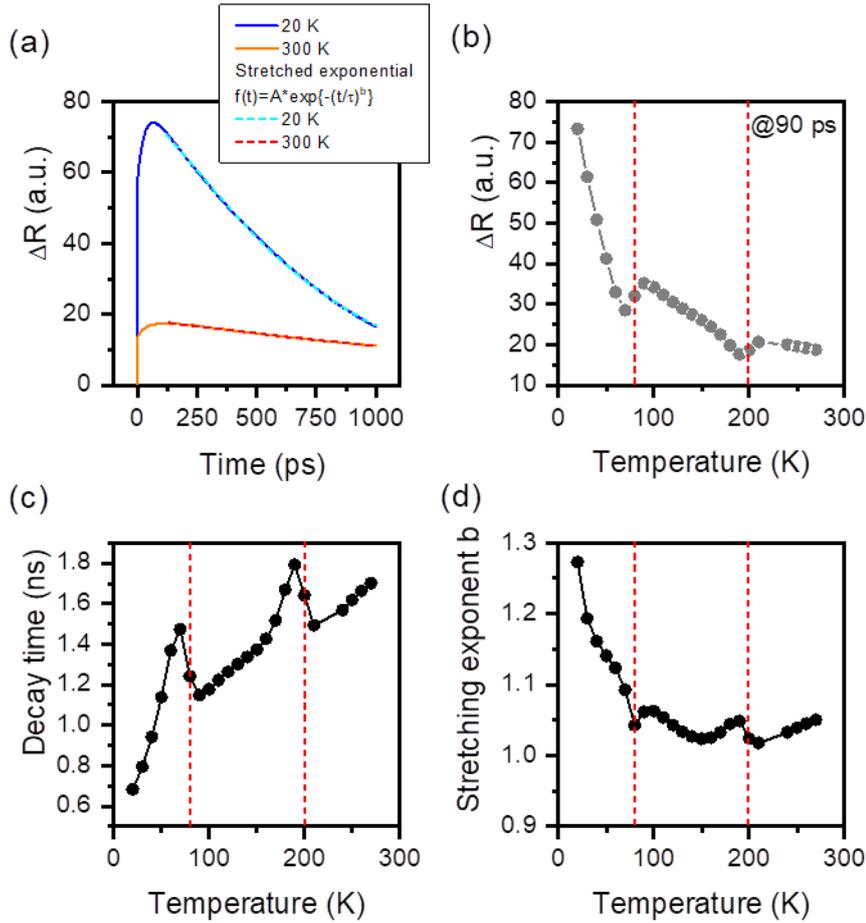

*FIG. S6*: Simulated pump probe reflectivity dynamics. (a) Emulated reflectivity signals $\Delta R(t)$ for two different base temperatures as indicated (solid lines), overlaid a fit of a stretched exponential function (dashed line). (b) Temperature dependence of $\Delta R(90\ ps)$. (c) and (d) depict the temperature dependence of the thermal decay time and of the stretching exponent b. Red vertical dashed lines indicate the phase transition temperatures.

Fig. S6 summarizes the results from the simulations. Typical reflectivity curves, emulated from simulations at base temperatures of 20 K and 300 K are shown in Fig. S6 (a). One can clearly see that, the relaxation towards equilibrium is much faster at low temperatures than at 300 K. Also, the maximum amplitudes are much larger, reflecting the much smaller specific heat of both the film and the substrate at low temperature. In Fig. S6 (b), the reflectivity signal 90 ps after the initial excitation is depicted as a function of the base temperature. Similar to the experimental findings, one sees an overall decrease of the reflectivity with increasing temperature. Around the phase transition temperature, anomalies appear, as explained in the main article.

Furthermore, the emulated reflectivity curves were fitted by a stretched exponential function $f(t) \sim e^{-\left(\frac{t}{\tau}\right)^b}$, yielding the temperature dependence of the decay time $\tau$ and of the stretching parameter $b$ shown in Fig S6 (c) and (d). Both show a monotonous behavior for temperatures away from phase transitions, and anomalies at the critical temperatures. Note that, quantitatively similar results can be obtained from an even simpler one-temperature model, because all subsystems are equilibrated anyway on the time scales of interest (100 ps to 1 ns). In such a one-temperature model, all contributions to the specific heat and the thermal conductivity are summed up.

In contrast to the real experiments, the simulations show an increase of the decay time from about 0.7 ns to 1.7 ns between 20 K and 300 K. We infer that, in particular at low temperatures, a long-living excitation is present in the experiments, whose live time exceeds the thermal decay time. A more realistic modelling of the manganite should incorporate this non-thermal process. Simply decreasing the electron-phonon and electron-spin coupling constant is not sufficient. Instead of electrons, a polaronic subsystem should be considered. The spatio-temporal diffusion of polarons and decay into phonons can maybe be described within the framework of a two-temperature model as derived for semiconductors.

## VI. Details on Theoretical simulation of the orbital ordering transition
### A. Heating and cooling protocol

The heating simulations started from the ground state described in the main text. The oxygen atoms received an initial random velocity distribution according to the target temperature of each simulation, followed by 2.4 ps of equilibration (24 thermostat cycles) with the thermostat at that temperature. Afterwards the Jahn-Teller structure factor was averaged for the next 12 ps. Starting from the 400 K heating simulation, the first set of cooling simulations reduced the target temperature of the thermostat to values between 20 K and 400 K. Again 2.4 ps of equilibration and 12 ps of averaging was used. A direct cooling from 400 K to low temperatures freezes the system in metastable states, as the cooling happens within one cycle of the thermostat of $P_{TS}$ =100 fs. Therefore, a second set of cooling simulations continued from the 200 K cooling calculations, again with the same equilibration and averaging times. The simulations recovered the initial orbital order of the ground state, although, in some cases, with the hole-polaron clusters distributed over several planes and thus with a slightly reduced order parameter. These more disordered metastable states are close in energy to the ground state, as explained in the main text. Heating simulations starting from such a metastable state led to the same transition temperature as the calculations shown in this paper.

### B. Orbital-order transition and Néel transition

In our simulation, the dominant $A$-type antiferromagnetic order breaks down simultaneously with the orbital order. In some simulation runs, we observe a small increase in the ferromagnetic moment during the transition. Above the orbital-order transition the system is paramagnetic. This is inconsistent with experiment where the Néel transition occurs at around 70 K and the system becomes paramagnetic around 80-130 K [2,9]. Both temperatures are substantially lower than in our simulations. This failure might originate from the adiabatic description of the spin dynamics in our simulation. In our current framework, we cannot yet verify whether the ferromagnetic moment is related to another magnetic phase transition.

## VII. Temperature dependent XRD studies: double-peak structure

The temperature dependence of the lattice parameters was deduced from the peak positions of the (200), (020) and (004) reflections. Fig. S7 shows the shift of these peaks with temperature for PCMO x=0 ((a), (b), (c)). For comparison, the shift of (004) PCMO x=0.1 peak is also shown (d).

For both compositions, the reflections reveal a double-peak shape, clearly visible in the shoulder-like feature of the (220) and (004) reflections ((c), (d)). For a coarse estimation of the volume fractions, the (004) reflections were fitted by two Gaussian peaks. Figure S8 shows such fits for the PCMO x=0.1 data in two different temperature ranges (230 – 300 K and 83 – 150 K). The data points at different temperatures were normalized to position and maximum height of the (004) reflection. The Gaussian peaks are practically unaffected by the temperature range and the peak areas correspond to a volume fraction of about 10% of the right-handed peak.

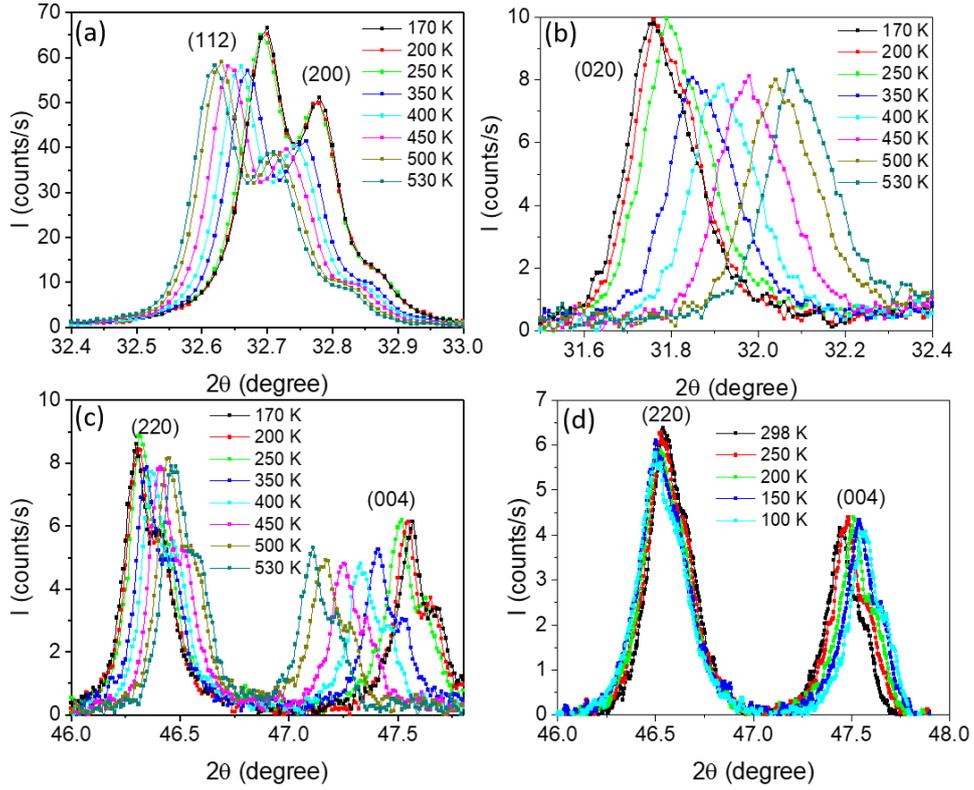

**FIG. S7**: (a), (b), (c) poly-crystalline PCMO x=0 bulk sample: Temperature dependence of angular position and shape of (a) (200) reflection, (b) (020) reflection and (c) (004) reflection. d) PCMO x=0.1 bulk sample, angular position and shape of (004) reflection.

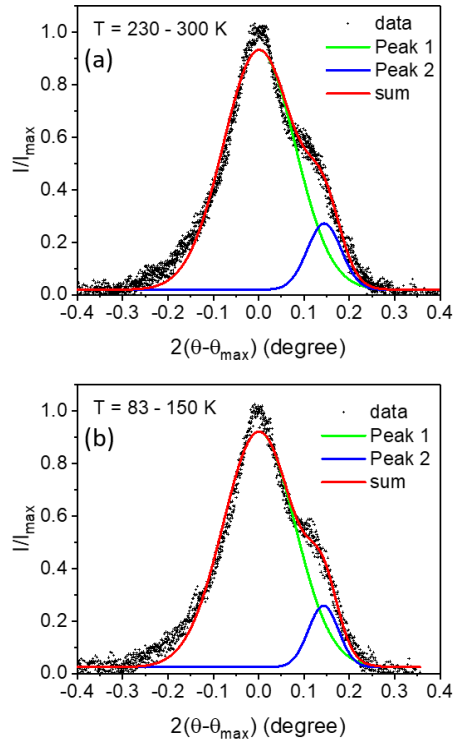

**FIG. S8**: (004) reflection PCMO x=0.1 bulk sample. The experimental data points are normalized to angular position and maximum intensity of the reflections at the respective temperature. (a) Data points in the temperature range from 230 to 300 K. (b) Data points in the temperature range from 83 to 150 K. The solid green and blue lines correspond to fit with two gaussian distributions (peak 1 and 2). The red line is the sum of both peaks.

## VIII. Extraction of the octahedral rotation axis from Rietveld refined data

As Rietveld refined positions due to experimental noise do not necessarily satisfy the condition of ideal $MnO_6$ octahedra, i.e. no distortion of the right angles, the rotation axis $(n_1, n_2, n_3)$ resp. angle $\theta$ are defined as minima of

$$(5) \quad \min_{n_1,n_2,n_3,\theta} \sum_{i=1}^{6} \|d_i \cdot \boldsymbol{R}(n_1,n_2,n_3,\theta)\vec{e}_i - \vec{o}_i\|^2 \quad \text{subject to } n_1^2 + n_2^2 + n_3^2 = 1$$

Here, $\vec{o}_i$ are the refined positions of the six oxygen atoms with respect to Mn, $d_i$ the experimentally found bonding lengths, and $\vec{e}_i$ unit vectors pointing to the position of the corresponding oxygen atom prior any rotation, i.e. along the Cartesian axes. $\boldsymbol{R}(n_1, n_2, n_3, \theta)$ denotes a three-dimensional rotation matrix with rotation axis $(n_1, n_2, n_3)$ and angle $\theta$. Noteworthy, the extracted values from [2,10] show little variations with doping and temperature justifying their fixation. In detail, solely $n_2 = 0.78$ is fixed, as the same geometrical consideration of a corner-shared octahedral network with Pbnm symmetry leading to eq.s (7a)-(7c) in the main text yields the following condition:

$$(6) \quad 0 = (d_l + d_s) \cdot (n_3 n_2 (1 - \cos\theta) + n_1 \sin\theta) + (d_l - d_s) \cdot (n_3 n_1 (1 - \cos\theta) - n_2 \sin\theta)$$

Combined with the normalization condition of $(n_1, n_2, n_3)$, fixing $n_2$ determines the rotation axis fully for given values $(d_l, d_s, \theta)$.